\documentclass[aps,prd,10pt,twocolumn,superscriptaddress,nofootinbib,floatfix]{revtex4-2}
\usepackage[T1]{fontenc}
\usepackage[utf8]{inputenc}
\usepackage{amsmath,amssymb,amsfonts,mathtools}
\usepackage{physics}
\usepackage{upgreek}
\usepackage{bm}
\usepackage{graphicx}
\usepackage{xcolor}
\usepackage{tikz}
\usetikzlibrary{arrows.meta,calc,positioning,decorations.pathreplacing}
\usepackage[colorlinks=true,linkcolor=blue,citecolor=magenta,urlcolor=cyan]{hyperref}

\def\beq{\begin{equation}}
\def\eeq{\end{equation}}

\newcommand{\ii}{\mathrm{i}}
\newcommand{\ee}{\mathrm{e}}

\newcommand{\conn}{\mathrm{conn}}

\newcommand{\disc}{\mathrm{disc}}
\DeclareMathOperator{\erfc}{erfc}
\providecommand{\erf}{\operatorname{erf}}
\newcommand{\Neg}{\mathcal{N}}
\newcommand{\NegRate}{\mathcal{R}}
\newcommand{\Psurv}{\mathcal{P}}
\newcommand{\AdS}{\mathrm{AdS}}
\newcommand{\avg}[1]{\langle #1 \rangle}
\newcommand{\Cplx}{\mathcal{C}}
\newcommand{\Nnorm}{\widetilde{\mathcal{N}}}

\begin{document}

\title{Wigner negativity of the Hartle--Hawking state in the length basis of JT gravity}

\author{Ritam Basu}
\email{ritam.basu@theory.tifr.res.in}
\affiliation{Department of Theoretical Physics, Tata Institute of Fundamental Research,\\
1 Homi Bhabha Road, Mumbai 400005, India}

\begin{abstract}
The Wigner function, in general, takes on negative values, and the amount
of negativity in the Wigner function gives an operationally meaningful
measure of the complexity of simulating the quantum state on a classical
computer. In this paper, we study the growth of Wigner negativity of the
Hartle--Hawking state of Jackiw--Teitelboim gravity under time evolution.
We work in the gravitational length basis, at genus zero, and at high
temperature, $\beta\ll1$. Our main analytic result is simple: to leading
order in $\beta$ the state is a Gaussian wavepacket of minimum
uncertainty, sitting at rest at the turning point of the Liouville wall.
Its Wigner function is therefore positive, and the negativity is
$1+\delta(\beta,t)$, where the correction $\delta$ is smaller than any
power of $\beta$. We show that $\delta$ is an even function of time, so
there is no linear growth at $t=0$, and that once the packet has
reflected off the wall its leading-order evolution is a Clifford shear,
which cannot change the negativity. We also numerically study the time
evolution of the Wigner function and its negativity. We observe that
$\delta$ stays below $10^{-11}$ at $t=0$ and through the reflection,
then rises slowly over a few tens of $\beta/\pi$, and then approaches a
late-time value consistent with
$\delta_\infty(\beta)\simeq0.08\,\ee^{-4.4/\beta}$ for $0.25\le\beta\le1$:
the lower the temperature, the larger the late-time negativity. The exact survival amplitude $Z(\beta+\ii t)/Z(\beta)$ fixes the spread complexity
to order $t^4$, and it also fixes the seed-normalised Wigner diagnostic
of our earlier work~\cite{Basu:2026inf,Basu:2026spp}.  We take this as evidence that the length basis is
ideally suited for a dual, semi-classical description of the dynamics at
genus zero. Beyond genus zero the length basis is overcomplete, and we
make no statement about finite $\ee^{S_0}$.
\end{abstract}

\maketitle

\section{Introduction and summary}
\label{sec:intro}

In general, given a classical phase space, there is a more-or-less standard
procedure one can follow to quantize the theory. But given a quantum theory,
it is not always clear whether there exists an underlying classical
description which is ``useful''. For instance, in the AdS/CFT correspondence,
the large $N$ boundary conformal field theory is a quantum mechanical system
at strong coupling, so the underlying classical description in terms of
classical boundary fields is far from being useful. However, it is a highly
non-trivial fact that a new useful classical description emerges in the form
of the bulk gravitational theory, with the bulk $\hbar\sim 1/N^2$. Given a
strongly coupled quantum system, it would be nice to have a general strategy
to identify when it admits a dual, semi-classical description, particularly
one involving gravity.

In order to state the result, we need to introduce the notion of the Wigner
function~\cite{Wigner:1932eb}. The Wigner function $W_\psi(q,p)$ for a state
$\psi$ is a quasi-probability distribution, which gives a representation of
the quantum state in phase space, parametrized here by $(q,p)$. This should
be contrasted against the wavefunction $\psi(q)$ of the state, whose
absolute square gives a probability distribution over a half-dimensional
(Lagrangian) submanifold, parametrized here by $q$. Importantly, the Wigner
function is not a true probability distribution because it can take on
negative values. Nevertheless, it is a natural quantum analog of a classical
probability distribution in phase space; for instance, it is real and unit
normalized, and under time evolution follows a quantum generalization of the
classical Liouville's equation for the time evolution of phase space
densities. When the Wigner function for a state is everywhere non-negative,
it acts as a genuine probability distribution, and one might roughly expect
such states to be ``classical''~\cite{Kenfack:2004}. Hudson's
theorem~\cite{Hudson:1974} shows that the only pure states for which the
Wigner function is everywhere positive are Gaussian states. In the
discrete setting of an odd-dimensional Hilbert space with a chosen
computational basis, the pure states with a non-negative Wigner function
are the stabilizer states~\cite{Gross:2006}, the discrete counterpart of
the Gaussian states. The negativity of the Wigner
function turns out to be an operationally meaningful measure (i.e., a
monotone under stabilizer operations) of the non-stabilizer content of a
state. Much like entanglement plays the role of a resource in the theory of
quantum communication, the negativity of the Wigner function plays the role
of a resource in the theory of quantum computation using stabilizer
circuits~\cite{Veitch:2012,Veitch:2013}. Indeed, Wigner negativity (or
``non-stabilizerness'') is an important feature of various interesting
physical systems including conformal field theories, condensed matter
systems, topological field theories, and chaotic quantum
systems~\cite{Goto:2021,Leone:2021,Leone:2022}. Interestingly, one can
generalize this argument to the case where circuit elements are not Wigner
positive, and one finds that the more the Wigner negativity in the circuit,
the less efficient the classical
simulation~\cite{Mari:2012,Pashayan:2015,Howard:2014}. The quantity
computed in the body of this paper is the continuous-variable Wigner
negativity in a canonical pair $(x,p)$; in that setting the analogous
operational statement is that states and operations with positive Wigner
functions admit efficient classical simulation~\cite{Mari:2012}, while the
relation between the continuous and the finite-dimensional notions at
finite $\ee^{S_0}$ is left open here.

Incidentally, the Krylov basis makes an interesting appearance in a certain
model of two dimensional quantum gravity called the double-scaled SYK
(DSSYK) model~\cite{Cotler:2016fpe,Lin:2022rbf}. In this correspondence, the
Krylov basis in the DSSYK model turns into the ``chord-length'' basis in the
dual description~\cite{Lin:2022rbf,Rabinovici:2023yex}. In the JT gravity
limit, the chord length precisely corresponds to the length of the bulk
wormhole, and furthermore, the growth of the spread complexity
of~\cite{Balasubramanian:2022tpr} corresponds to the growth of the length
(what would have been volume in higher dimensions) of the wormhole interior,
in line with the ``Complexity = Volume''
conjecture~\cite{Susskind:2014rva,Stanford:2014jda}. From our point of view,
it is natural to regard the Krylov basis as a non-perturbative
generalization of the geometric length basis of
gravity~\cite{Lin:2022rbf,Rabinovici:2023yex}. One may say that gravity
provides the most efficient set of variables to describe aspects of the
strong-coupling dynamics in the dual quantum mechanical theory at large $N$.

In order to further explore this proposal for ``gravity as efficient
classical implementation of quantum computation'', it would be helpful to
study Wigner negativity and its growth in models such as DSSYK or the
Saad-Shenker-Stanford matrix model~\cite{Saad:2019lba}. It would also be
useful to generalize our results to the continuum setting. In
\cite{Basu:2024vem}, we focused on the time-dependence of the KW-negativity
in random matrix theory with the initial state being a generic pure state at
infinite temperature. We can include finite temperature by taking the
initial state to be of the form $|\psi_0\rangle=Z_\beta^{-1/2}\ee^{-\beta
H/2}|0\rangle$, where $\beta$ is the inverse temperature. However, as the
temperature is lowered, the saturation value of the negativity decreases
(although it remains proportional to $\sqrt D$). It would be interesting to
explore the temperature dependence of the negativity in more detail. A
significant obstacle is that the Wigner function and its negativity are
difficult to compute analytically, but as a first step towards this goal,
here we study the time evolution of the Wigner function of the
Hartle--Hawking (HH) state~\cite{Hartle:1983ai} and its negativity in JT
gravity~\cite{Teitelboim:1983ux,Jackiw:1984je,Almheiri:2014cka,Maldacena:2016upp}
(see~\cite{Mertens:2022irh} for a review).

JT gravity in asymptotically $\AdS_2$ spacetimes is dual to a quantum
mechanical theory on the asymptotic boundary -- the path integral of JT
gravity computes ensemble averaged quantities from the boundary point of
view~\cite{Saad:2019lba}, where the averaging is over an ensemble of
Hamiltonians. The chord-number basis is the natural generalization of the
gravitational length basis (i.e., a basis for the gravitational Hilbert
space labelled by the length of the bulk wormhole) in JT
gravity~\cite{Maldacena:2016upp,Yang:2018gdb,Harlow:2018tqv,Iliesiu:2024cnh}.
The survival probability is precisely the spectral form factor (SFF)
$\avg{|Z(\beta+\ii t)|^2}$~\cite{Cotler:2016fpe,Garcia-Garcia:2016mno,Saad:2019lba},
whose ramp is the connected double-trumpet
wormhole~\cite{Saad:2019lba,Saad:2019pqd}; it is known that the sub-leading
corrections to this formula become important at a time of order
$t\sim O(\ee^{S_0/2})$. The central result of this paper is that at
genus zero, in the high-temperature regime $\beta\ll1$ and to leading
order in $\beta$, the HH state is a Gaussian wavepacket in the length
basis, so that its negativity is $1+\delta(\beta,t)$ with $\delta$ beyond
all orders in $\beta$, even in $t$ and with no linear rise at $t=0$; a
direct numerical evaluation of $\delta$ (Sec.~\ref{sec:numerics}) finds
it below $10^{-11}$ through the reflection and frozen afterwards at a
value compatible with $\ee^{-c/\beta}$, $c\simeq4.4$. Any exponentially
large value can only come from the discreteness of the spectrum, where
the length basis ceases to be a basis (Sec.~\ref{sec:beyond}); we make no
statement about it. The restriction to $\beta\ll1$ is essential: the
low-temperature regime, where the HH state is dominated by near-threshold
energies, is not treated here.

\subsection{Results}
\label{sec:results}

Our main results are as follows:
\begin{itemize}
\item \textbf{The Hartle--Hawking packet.} To leading order in $\beta$
the HH state in the length basis is a minimum-uncertainty Gaussian at rest
at the classical turning point, Eq.~\eqref{eq:HHgauss}, with
$\langle x\rangle=2\log(2\beta/\pi)$, $\sigma_x^2=2\beta/\pi^2$ and
$\langle p^2\rangle=\pi^2/8\beta$; the leading saddle-point prefactor is
consistent with unit normalisation. Its Wigner function is positive, and
time-reversal symmetry makes $\Neg(t)$ even in $t$, so there is no linear
term at $t=0$ [Eq.~\eqref{eq:Nparity}].
\item \textbf{Negativity.} $\Neg(\beta,t)=1+\delta(\beta,t)$ at genus
zero, with $\delta>0$ by Hudson's theorem but beyond all orders in
$\beta$ (Sec.~\ref{sec:negativity}). We also numerically study the time
evolution of the Wigner function and its negativity
(Sec.~\ref{sec:numerics}, Fig.~\ref{fig:neg}). We observe that $\delta$
is below $10^{-11}$ at $t=0$ and through the main reflection for
$0.25\le\beta\le1$, then rises gradually, as the low-momentum tail of the
packet reflects, and saturates at
$\delta_\infty(\beta)\simeq0.08\,\ee^{-4.4/\beta}$, from
$2.2\times10^{-9}$ at $\beta=0.25$ to $1.1\times10^{-3}$ at $\beta=1$.
\item \textbf{Freezing.} Any quantum evolution which starts from a
Gaussian state and involves only Clifford gates will keep the state within
the restricted class of Gaussian states. Once the packet has left the wall
its evolution is, up to corrections exponentially small in $1/\beta$,
generated by a Hamiltonian quadratic in the canonical variables, under
which $\int|W|$ is exactly invariant [Eqs.~\eqref{eq:covariance}
and~\eqref{eq:freeshear}]; the numerical $\delta(t)$ approaches a frozen
value, reached to within $5\%$ at the longest time computed
($t=256\,\beta/\pi$ at $\beta=1$).
\item \textbf{SFF.} On the same footing, the SFF displays a ramp at times
of $O(\ee^{S_0/2})$~\cite{Saad:2019lba,Okuyama:2019xbv,Saad:2021rcu}
because its disk term decays as $t^{-3}$; the disk term of the negativity
does not decay, so the two-boundary wormhole corrects the averaged Wigner
function only at relative order $t\,\ee^{-2S_0}$ (Sec.~\ref{sec:beyond}).
This is a remark about the mean Wigner function: beyond genus zero the
length basis is overcomplete and $\int|W|$ is not defined, so no
statement is made about the negativity itself at finite $\ee^{S_0}$.
\item \textbf{Spread complexity.} The exact survival amplitude
$S(t)=Z_{\beta+\ii t}/Z_\beta$ shows that the survival probability decays
away from one as a function of time, with
$\Cplx(t)=\sigma_E^2t^2-2\pi^4t^4/[\beta^5(4\pi^2+3\beta)]+O(t^6)$ exactly at
early times, Eq.~\eqref{eq:C4}, and Lanczos coefficients
$a_n=2n/\beta+4\sqrt n/\beta^{3/2}+O(1)$,
$b_n=n/\beta+2\sqrt n/\beta^{3/2}+O(1)$; the late-time growth law is
not derived (Sec.~\ref{sec:spread}).
\item \textbf{Normalised negativity.} The seed-normalised negativity
of~\cite{Basu:2026inf} is exactly $\Nnorm=\Neg/\Psurv$ with
$\Psurv=|S(t)|^2$, so at genus zero it is the inverse of the exact
survival probability up to $\delta$, Eq.~\eqref{eq:Nnormdisk}; its $t^3$
growth and $3/t$ rate are properties of the survival probability
(Sec.~\ref{sec:normneg}).
\end{itemize}

\subsection{Disk versus double trumpet}
\label{sec:intro-time}

One key point is that even at small times, the ensemble average of the
negativity of the Wigner function is not the same as the negativity of the
averaged Wigner function. Furthermore, the negativity involves a
non-analytic function (namely, the absolute value of the Wigner function),
so a small correction to the mean $\avg{W}$ does not control $\avg{|W|}$.
The survival probability is precisely the spectral form factor. The leading
contribution to this, of course, comes from the disconnected answer (i.e.,
the product of the averages); the ramp is the connected contribution, the
double trumpet, which is $\ee^{-2S_0}$-suppressed and becomes visible only
because the disconnected answer decays as a power of $t$. For the
negativity of the HH state the disconnected answer is the disk Wigner
function of a Gaussian packet, positive and, at leading order, time
independent, which does not decay, so the same double trumpet corrects
the averaged Wigner function only at relative order $t\,\ee^{-2S_0}$.
Section~\ref{sec:beyond} states this, and also why it is where the
genus-zero analysis stops: at finite $\ee^{S_0}$ the length basis is
overcomplete and the continuum negativity is not defined.

\subsection{Relation to the Krylov--Wigner programme}
\label{sec:intro-programme}

In~\cite{Basu:2024vem}, a perturbative argument was given to show that the
Krylov basis minimizes the early time growth of Wigner negativity in the
large-$D$ limit (see~\cite{Balasubramanian:2026wkv,Basu:2026inf,Basu:2025stb,Bhattacharya:2026bda,Bhattacharya:2026yfm}
for further developments). Using tools from random matrix theory,
in~\cite{Basu:2025kjr} we showed that for a generic choice of basis, the
Wigner negativity for a classical initial state becomes exponentially large
in an $O(1)$ amount of time evolution, with an explicit formula in terms of
the survival amplitude, while in the Krylov basis the negativity grows at
most as a power law, and becomes exponentially large only at exponential
times. The KW negativity of~\cite{Basu:2024vem} thus shows a much more
gradual rise than a generic basis; in the continuum length basis of JT this
becomes, at genus zero, no rise at all beyond the exponentially small
$\delta$ (Sec.~\ref{sec:diskN}). The continuous-variable negativity in
$(x,p)$ and the discrete negativity in the Krylov chain are different
objects, and we do not relate them quantitatively.

\subsection{Relation to earlier diagnostics, and plan}
\label{sec:intro-plan}

The Wigner negativity gives a basis-dependent, operational measure of the
complexity of simulating the time evolution of a quantum state on a
classical computer. It is naturally related to other measures of complexity,
such as the spread complexity, and provides a concrete notion of
classicality for abstract quantum systems. Wigner negativity gives a lower
bound on a certain entropic measure of spread complexity in the discrete
setting~\cite{Basu:2024vem}. We should emphasize that by complexity, we do
not mean circuit
complexity~\cite{Susskind:2014rva,Stanford:2014jda,Brown:2015bva,Belin:2021bga},
although it would be interesting to study the connection to circuit
complexity further; nor do we address the late-time saturation expected of
complexity measures, which lies beyond genus zero.

The plan for the paper is as follows: in Sec.~\ref{sec:setup} we review
some background material about the Wigner function and fix conventions. In
Sec.~\ref{sec:negativity}, we derive the Gaussian form of the HH packet,
its reflection and its negativity, evaluate $\delta$ numerically, and
record what the double trumpet does and does not say.
Sections~\ref{sec:spread} and~\ref{sec:normneg} present the spread
complexity and the seed-normalised negativity, Sec.~\ref{sec:master}
collects the results with their status, and Sec.~\ref{sec:disc} discusses
open questions. Appendices~\ref{app:bessel}--\ref{app:cumulants} contain
the details of the derivations.

\section{Setup and conventions}
\label{sec:setup}

\subsection{The Hartle--Hawking state in the length basis and the Liouville wall}
\label{sec:HHstate}

The chord-number basis of the double-scaled SYK model, which is the Krylov
basis of its infinite-temperature thermofield double, discretises JT
gravity away from the low-energy
limit~\cite{Lin:2022rbf,Rabinovici:2023yex,Bhattacharya:2026yfm}. At low
energies the chord number becomes a continuous variable, the renormalised
length of the wormhole between the two boundaries of the eternal black
hole, and the chord Hamiltonian reduces to the Liouville Hamiltonian of
JT gravity~\cite{Bagrets:2016cdf,Harlow:2018tqv,Maldacena:2016upp,Yang:2018gdb,Kitaev:2018wpr,Bhattacharya:2026yfm}.
We work directly in this continuum length basis, with the renormalised
length $x$ as the canonical coordinate and $p$ its conjugate momentum (the
variable called $\ell$ in~\cite{Bhattacharya:2026yfm} is our $x$; we
reserve $\ell\equiv\ee^{x/2}$ for the auxiliary coordinate in which linear
growth of $x$ is exponential growth of $\ell$).

More precisely, JT gravity is dual to a double-scaled random matrix theory,
with the leading contribution to the density of states given
by~\cite{Stanford:2017thb,Mertens:2017mtv,Bagrets:2016cdf}
\beq
  \rho(E)=\frac{\ee^{S_0}}{2\pi^2}\,\sinh\!\big(2\pi\sqrt E\,\big),
  \qquad \rho_0(s)\equiv\sinh(2\pi s),
  \label{eq:dos}
\eeq
where energy eigenstates are labelled by $E=s^2$ with $s>0$; the density
is unbounded above, so it carries no normalisation of its own.\footnote{Equation~\eqref{eq:dos}
defines our $S_0$. The density of~\cite{Saad:2019lba} at their
$\gamma=\tfrac12$ carries $1/(4\pi^2)$ instead of $1/(2\pi^2)$, while the
trumpet and the double trumpet [Eq.~\eqref{eq:ramp}] are the same in both
conventions; hence $\ee^{S_0}$ here is one half of the $\ee^{S_0}$
of~\cite{Saad:2019lba}, and in their convention the dip
time~\eqref{eq:tdip} reads $\beta^{1/4}\ee^{S_0/2}/\sqrt2$.} We can regard $S_0$ as the entropy of
the higher-dimensional extremal black hole for which the JT theory is the
two-dimensional reduction, or it can be thought of as the ground state
entropy of the JT system.
All explicit results of this paper are at genus zero (the disk), where
$S_0$ does not appear: the only expansion parameter is $\beta$, which
plays the role of $\hbar$ in the phase-space description below. The
parameter $\ee^{-S_0}$ enters only in Sec.~\ref{sec:beyond}.

The fixed-energy wavefunctions in the length basis are modified Bessel
functions of imaginary
order~\cite{Yang:2018gdb,Mertens:2017mtv,Kitaev:2018wpr,Bagrets:2016cdf},
\beq
  \chi_E(x)=K_{2\ii s}\!\big(4\ee^{-x/2}\big),\qquad E=s^2,
  \label{eq:chidef}
\eeq
as follows from the exact quantisation of the boundary Schwarzian mode.
These wavefunctions obey a Liouville-type Wheeler--DeWitt equation. Let
$z\equiv 4\ee^{-x/2}$, so that $\dd z/\dd x=-z/2$; for $\chi(x)=K(z(x))$,
\begin{equation*}
  \begin{split}
  \frac{\dd^2\chi}{\dd x^2}
  &=\frac{z}{4}K'(z)+\frac{z^2}{4}K''(z)
  =\frac14\big[z^2K''+zK'\big]\\
  &=\frac14\big(z^2-4s^2\big)\chi,
  \end{split}
\end{equation*}
where the last step used the modified Bessel equation of order $\nu=2\ii s$,
$z^2K''+zK'=(z^2+\nu^2)K=(z^2-4s^2)K$. Hence, for every $E>0$,
\beq
  -\,\chi_E''(x)+4\,\ee^{-x}\,\chi_E(x)=E\,\chi_E(x),
  \label{eq:liouville}
\eeq
i.e.\ the $\chi_E$ are the delta-normalisable eigenfunctions of the
Liouville Hamiltonian $H_{\rm WdW}=p^2+4\ee^{-x}$. The classical turning
point, where the local momentum $k(x)=\sqrt{E-4\ee^{-x}}$ vanishes, sits at
\beq
  x_{\rm turn}(s)=\log\frac{4}{E}=2\log\frac{2}{s},
  \label{eq:xturn}
\eeq
and $k(x)\to s$ for $x\gg x_{\rm turn}$. The potential $V(x)=4\ee^{-x}$ is
the Liouville wall: impenetrable as $x\to-\infty$, and decaying
exponentially in $x$ (as $4/\ell^2$ in $\ell=\ee^{x/2}$) on the allowed
side. This last property is what will make the disk-level negativity
freeze, asymptotically, once the reflected packet has left the wall
(Sec.~\ref{sec:freezing}).

We can include finite temperature by taking the initial state to be of the
form
\beq
  \psi_\beta(x)=\frac{\sqrt{2}}{\pi\sqrt{Z_\beta}}
  \int_0^\infty\! 2s\,\dd s\;\rho_0(s)\,\ee^{-\frac{\beta}{2}s^2}\,
  K_{2\ii s}\!\big(4\ee^{-x/2}\big),
  \label{eq:psi}
\eeq
where $\beta$ is the inverse temperature, with $E=s^2$, measure
$\rho(E)\,\dd E=\rho_0(s)\,2s\,\dd s$, and the normalisation
$\langle\psi_\beta|\psi_\beta\rangle=1$ (Appendix~\ref{app:norm}).
The factor $\ee^{-\beta s^2/2}=\ee^{-\beta E/2}$ in~\eqref{eq:psi} is the
half-circle Euclidean weight: $|\psi_\beta\rangle\propto\ee^{-\beta H/2}|\,\cdot\,\rangle$
prepared from the boundary state at zero length, and $Z_\beta$ is the norm
of this unnormalised state, i.e.\ the overlap of two half-circle
Hartle--Hawking states.

\subsection{The exact Schwarzian partition function}
\label{sec:Z}

The reduced partition function can be evaluated in closed form: for all
$\beta>0$,
\beq
  Z_\beta\;\equiv\;\int_0^\infty 2s\,\sinh(2\pi s)\,\ee^{-\beta s^2}\,\dd s
  \;=\;\frac{\pi^{3/2}}{\beta^{3/2}}\,\ee^{\pi^2/\beta}.
  \label{eq:Zdef}
\eeq
To see this, write $\sinh(2\pi s)=\tfrac12(\ee^{2\pi s}-\ee^{-2\pi s})$, so
that
\beq
  Z_\beta=I_+-I_-,
  \qquad
  I_\pm\equiv\int_0^\infty s\,\ee^{\pm2\pi s-\beta s^2}\,\dd s .
  \label{eq:Zsplit}
\eeq
Completing the square,
$\pm2\pi s-\beta s^2=-\beta(s\mp\tfrac\pi\beta)^2+\tfrac{\pi^2}{\beta}$,
exhibits the saddle at $s=\pm\pi/\beta$ with common peak value
$\ee^{\pi^2/\beta}$. With $c\equiv\pi/\beta$, substituting $u=s-c$ in $I_+$
and $u=s+c$ in $I_-$ gives
\beq
  \begin{split}
  I_+&=\ee^{\pi^2/\beta}\int_{-c}^{\infty}(u+c)\,\ee^{-\beta u^2}\dd u,\\
  I_-&=\ee^{\pi^2/\beta}\int_{c}^{\infty}(u-c)\,\ee^{-\beta u^2}\dd u .
  \end{split}
  \label{eq:Ipm}
\eeq
The linear-weight pieces are exact derivatives,
$\int_{a}^{\infty}u\,\ee^{-\beta u^2}\dd u=\ee^{-\beta a^2}/2\beta$, equal
at $a=\pm c$, so they cancel identically in $I_+-I_-$. The Gaussian-weight
pieces add: using
$\int_{-c}^{\infty}=\int_{-\infty}^{\infty}-\int_{-\infty}^{-c}$ and
evenness, the two half-line tails cancel and only the full Gaussian
survives,
\beq
  I_+-I_-
  =\ee^{\pi^2/\beta}\,c\int_{-\infty}^{\infty}\ee^{-\beta u^2}\dd u
  =\ee^{\pi^2/\beta}\,\frac{\pi}{\beta}\sqrt{\frac{\pi}{\beta}} ,
  \label{eq:J0}
\eeq
which is~\eqref{eq:Zdef}, exact for all $\beta>0$ --- no high-temperature
approximation is made. [In error-function language the same step reads
$J_0^+=\tfrac12\sqrt{\pi/\beta}\,(1+\erf\sqrt\beta c)$,
$J_0^-=\tfrac12\sqrt{\pi/\beta}\,\erfc(\sqrt\beta c)$, combining through
$\erf+\erfc=1$.]

Restoring the prefactor of~\eqref{eq:dos},
\beq
  Z^{\rm JT}(\beta)=\frac{\ee^{S_0}}{2\pi^2}\,Z_\beta
  =\frac{\ee^{S_0}}{2\sqrt\pi}\,\beta^{-3/2}\,\ee^{\pi^2/\beta},
  \label{eq:ZJT}
\eeq
and from $\log Z_\beta=\tfrac32\log\pi-\tfrac32\log\beta+\pi^2/\beta$ the
exact thermal energy moments follow,
\beq
  \begin{split}
  \langle E\rangle_\beta&=-\partial_\beta\log Z_\beta
  =\frac{\pi^2}{\beta^2}+\frac{3}{2\beta},\\
  \langle E^2\rangle_\beta-\langle E\rangle_\beta^2&=\partial_\beta^2\log Z_\beta
  =\frac{2\pi^2}{\beta^3}+\frac{3}{2\beta^2}.
  \end{split}
  \label{eq:Emoments}
\eeq
The same peak controls the HH energy distribution
$\rho(E)\ee^{-\beta E}\propto\ee^{2\pi s-\beta s^2}$, sharply peaked at
\beq
  s_*=\frac{\pi}{\beta},\qquad E_*=\frac{\pi^2}{\beta^2},\qquad
  \text{width }\;\sigma_s=\frac{1}{\sqrt{2\beta}} ,
  \label{eq:saddle}
\eeq
consistent with the leading terms of~\eqref{eq:Emoments}:
$\langle E\rangle_\beta\simeq E_*$ and
$\mathrm{Var}(E)\simeq(2s_*\sigma_s)^2=2\pi^2/\beta^3$ at high temperature.
The semiclassical disk analysis is controlled at $\beta\ll1$, where
$s_*\gg1$. (The amplitude itself peaks at the same $s_*$ but with the wider
Gaussian width $\sigma_s^{\rm amp}=1/\sqrt\beta$, because the amplitude
carries $\ee^{-\beta s^2/2}$ rather than $\ee^{-\beta s^2}$.) Every overall
constant in $\psi_\beta$ cancels against the exact sum rule
$\int W\,\dd x\,\dd p=1$ imposed below, so the negativity is independent of
the normalisation of $Z_\beta$.

\paragraph{Lorentzian evolution.} The states of interest are
$\ee^{-zH}|\,\cdot\,\rangle$ with complex preparation parameter
$z=\tfrac{\beta}{2}+\ii t$: consider the time evolved state
$|\psi(t)\rangle=\ee^{-\ii Ht}|\psi_\beta\rangle$, which replaces
$\ee^{-\beta H/2}$ by $\ee^{-(\beta/2+\ii t)H}$, i.e.\ in~\eqref{eq:psi}
\beq
  \ee^{-\frac{\beta}{2}s^2}\;\to\;\ee^{-(\frac{\beta}{2}+\ii t)s^2},
  \qquad \psi_t(x)\equiv\langle x|\psi(t)\rangle .
  \label{eq:psit}
\eeq
Here $H$ is the Liouville Hamiltonian of~\eqref{eq:liouville}, $H=p^2+4\ee^{-x}$
with $E=s^2$, and $t$ is the boundary time: this single convention is used
throughout the paper, so that the survival amplitude of Sec.~\ref{sec:spread},
$\langle\psi_\beta|\psi(t)\rangle=Z_{\beta+\ii t}/Z_\beta$, and the
phase-space evolution of Sec.~\ref{sec:negativity} refer to the same $t$.
Hamilton's equations for $H$ are $\dot x=2p$, $\dot p=4\ee^{-x}$.
Evolution is unitary, so $\int\dd x\,|\psi_t(x)|^2=1$ for all $t$.

\subsection{The Wigner function: sum rules and exact invariances}
\label{sec:wig}

Given a quantum theory, how does one recover the underlying classical phase
space? The Wigner function approach is one way to proceed, where one
defines a quasi-probability distribution called the Wigner function, on a
phase space fixed by the choice of a computational basis:
\begin{align}
  W(x,p;t)&=\frac{1}{2\pi}\int_{-\infty}^{\infty}\!\dd y\;\ee^{\ii py}\,
  \psi^*_{t}\!\big(x+\tfrac{y}{2}\big)\,
  \psi_{t}\!\big(x-\tfrac{y}{2}\big),
  \label{eq:Wdef}\\[1mm]
  \Neg(t)&=\int\!\dd x\,\dd p\;|W(x,p;t)| .
  \label{eq:Ndef2}
\end{align}
It is the closest quantum analogue of a classical phase-space density,
and shares several of its properties:
\begin{enumerate}
\item The Wigner function is real and unit normalized:
$\int W_\psi\,\dd x\,\dd p=\|\psi\|^2$, since
$\int\dd p\,\ee^{\ii p y}=2\pi\delta(y)$ collapses the $y$ integral
in~\eqref{eq:Wdef}.
\item Integrating the Wigner function along one-direction gives the
probability density. For e.g., $\int\dd p\,W(x,p)=\langle x|\rho|x\rangle$.
\item The time evolution of the Wigner function is controlled by the Moyal
equation, which reduces to the Liouville equation in the $\hbar\to0$ limit
[Eq.~\eqref{eq:moyaleq} below].
\end{enumerate}
What it is not is a probability distribution, since it can be negative.
Inserting two copies of
the definition, the $p$ integral of $W_\varphi W_\psi$ sets $y'=-y$;
substituting $u=x+\tfrac y2$, $v=x-\tfrac y2$ (unit Jacobian) the integral
factorises into the Moyal identity
\beq
  \int W_\psi\,\dd x\,\dd p=\|\psi\|^2,
  \quad
  \int W_\varphi\,W_\psi\,\dd x\,\dd p
  =\frac{1}{2\pi}\big|\langle\varphi|\psi\rangle\big|^2 .
  \label{eq:moyal}
\eeq
In particular, the pure normalised state~\eqref{eq:psi} satisfies at any
time
\beq
  \int W\,\dd x\,\dd p=1,\qquad
  \int W^2\,\dd x\,\dd p=\frac{1}{2\pi}\,\Tr\rho_\psi^2=\frac{1}{2\pi},
  \label{eq:moments}
\eeq
and writing $W_-\equiv\max(-W,0)$ for the negative part,
$|W|=W+2W_-$ pointwise, so $\Neg=1+2\int W_-$.

But even if we were to start with a classical/Gaussian state, the positivity
of the Wigner function will not, in general, be preserved under time
evolution. This only happens for at most inhomogenous, quadratic (i.e.,
free) Hamiltonians. The space of such at most quadratic Hamiltonians
generates a group, which we may call the ``Clifford'' group. Let
$S:\mathbb R^2\to\mathbb R^2$ be an affine symplectic map,
$(x,p)\mapsto M(x,p)^{\rm T}+v$ with $\det M=1$, and $U_S$ the
corresponding metaplectic (Clifford) unitary, generated by Hamiltonians at
most quadratic in $(\hat x,\hat p)$. The Wigner function evolves by the
Moyal equation
\beq
  \begin{split}
  \partial_t W&=\{h,W\}_{\rm MB}
  =\{h,W\}_{\rm PB}\\
  &\quad+\sum_{n\ge1}\frac{(-1)^n}{4^n(2n+1)!}\,
  h\big(\overleftarrow{\partial_x}\overrightarrow{\partial_p}
  -\overleftarrow{\partial_p}\overrightarrow{\partial_x}\big)^{2n+1}W ,
  \end{split}
  \label{eq:moyaleq}
\eeq
with $h(x,p)$ the Weyl symbol of the generator. Every term in the sum
carries at least three derivatives of $h$, so for $h$ polynomial of degree
at most two they vanish identically and $W$ is transported along the
classical affine flow, $W_t(x,p)=W_0(S_t^{-1}(x,p))$. Since $S^{-1}$ has
unit Jacobian,
\beq
  \begin{split}
  W_{U_S\psi}(x,p)&=W_\psi\big(S^{-1}(x,p)\big),\\
  \Neg[U_S\psi]&=\Neg[\psi].
  \end{split}
  \label{eq:covariance}
\eeq
More generally, the Wigner function transforms covariantly under any
Clifford unitary, and so the negativity remains invariant under such
Clifford unitary transformations -- phase-space translations, rotations,
squeezing, and free evolution.

For a particle on a line the natural computational basis is the position
basis in which the Hamiltonian is a differential operator of second order;
here that basis is the length basis $x$ of~\eqref{eq:liouville}, with $p$
its conjugate momentum. A crucial point is that the Wigner function and
its negativity is defined with respect to a choice of computational basis.
Within the fixed canonical pair $(x,p)$ with measure $\dd x\,\dd p$, the
negativity is a well-defined single number at each time, exactly invariant
under the affine symplectic maps of~\eqref{eq:covariance}. It is not
invariant under the nonlinear point transformation $x=2\log\ell$ to the
auxiliary length $\ell$, since the Wigner function is not a scalar under
nonlinear canonical maps; we therefore compute throughout in the canonical
$(x,p)$ frame and make no claim about the $(\ell,p_\ell)$ frame.

\section{The Wigner negativity of the Hartle--Hawking state}
\label{sec:negativity}

Having discussed the importance of Wigner negativity in the theory of
quantum computation, our aim now is to study the behavior of the Wigner
function and its negativity under time evolution for the Hartle--Hawking
state. Specifically we will be interested in the degree to which the time
evolution of a classical initial state remains semiclassical with respect
to this measure.

\subsection{The Hartle--Hawking packet at the turning point}
\label{sec:HHpacket}

The first question is what the state~\eqref{eq:psi} looks like in the
length basis at $t=0$. Using the integral representation
$K_{2\ii s}(z)=\int_0^\infty\dd v\,\ee^{-z\cosh v}\cos(2sv)$ and carrying
out the Gaussian $s$ integral (Appendix~\ref{app:packet}), the HH
wavefunction becomes a single integral,
\beq
  \begin{split}
  \psi_\beta(x)&=\frac{2}{\sqrt{\pi Z_\beta}}\,\frac{\ee^{2\pi^2/\beta}}{\beta^{3/2}}\,
  \mathrm{Re}\!\int_{-\infty}^{\infty}\!\dd v\;(\pi+\ii v)\\
  &\qquad\times\exp\!\Big[-4\ee^{-x/2}\cosh v-\frac{2v^2}{\beta}+\frac{4\pi\ii v}{\beta}\Big],
  \end{split}
  \label{eq:psi-v}
\eeq
in which $\beta$ plays the role of $\hbar$: every term in the exponent is
$O(1/\beta)$ once $x$ is measured from the classical turning point
$x_{\rm turn}(s_*)=2\log(2\beta/\pi)$ of the peak energy~\eqref{eq:saddle}.
The saddle point of~\eqref{eq:psi-v} lies on the imaginary $v$ axis, and
its evaluation (Appendix~\ref{app:packet}) shows that the exponent is real
and is minimised exactly at $x=x_{\rm turn}(s_*)$, with curvature
$\pi^2/4$. The HH state is therefore, to leading order in the
semiclassical parameter $\beta$, a real Gaussian wavepacket at rest at the
turning point,
\beq
  \boxed{\;
  \psi_\beta(x)\simeq\Big(\frac{\pi}{4\beta}\Big)^{1/4}
  \exp\!\Big[-\frac{\pi^2}{8\beta}\big(x-x_{\rm turn}(s_*)\big)^2\Big],\;}
  \label{eq:HHgauss}
\eeq
with
\beq
  \begin{split}
  &\langle x\rangle=2\log\frac{2\beta}{\pi},\qquad
  \sigma_x^2=\frac{2\beta}{\pi^2},\\
  &\langle p\rangle=0,\qquad\qquad\quad\;\;
  \langle p^2\rangle=\frac{\pi^2}{8\beta},
  \end{split}
  \label{eq:HHmoments}
\eeq
a minimum-uncertainty packet, $\sigma_x\sigma_p=\tfrac12$. The prefactor
in~\eqref{eq:HHgauss} is not adjusted by hand: the leading saddle-point
prefactor of~\eqref{eq:psi-v} is consistent with the unit normalisation of
Appendix~\ref{app:norm}, which is a non-trivial check of the computation
(the Gaussian itself is of course not the exact wavefunction). The corrections to~\eqref{eq:HHgauss} are relatively
$O(\sqrt\beta)$ and odd in $x-x_{\rm turn}$, i.e.\ a small skewness.

Two features of~\eqref{eq:HHmoments} deserve comment. First, the packet
has zero mean momentum although its mean energy is
$\langle E\rangle_\beta\simeq s_*^2=\pi^2/\beta^2$: at $t=0$ the energy is
entirely potential, $\langle4\ee^{-x}\rangle\simeq s_*^2$, and
$\langle p^2\rangle=O(1/\beta)$ is only the zero-point spread of the
packet. The HH state at $t=0$ is not a superposition of an incoming and an
outgoing wave, but a single lump sitting at the wall; this is the
Lorentzian time-symmetric point of the classical trajectory, at which the
Einstein--Rosen bridge has its minimal length. Second, the Wigner function
of~\eqref{eq:HHgauss} is the positive Gaussian
\beq
  W(x,p;0)\simeq\frac1\pi\,
  \exp\!\Big[-\frac{\pi^2}{4\beta}\big(x-x_{\rm turn}\big)^2-\frac{4\beta}{\pi^2}\,p^2\Big]
  \;\ge\;0 ,
  \label{eq:HHwigner}
\eeq
so that $\Neg(0)=1$ at this order. When the Wigner function for a state is
everywhere non-negative, it acts as a genuine probability distribution, and
one might roughly expect such states to be ``classical'': the HH state is
Wigner-classical with respect to the length basis in this sense at $t=0$,
to leading order in $\beta$; we stress that Wigner positivity is a
basis-dependent property and does not by itself define a basis-independent
notion of classical dynamics.

The classical trajectory launched from~\eqref{eq:HHmoments} is the
solution of $\dot x=2p$, $\dot p=4\ee^{-x}$ [Hamilton's equations for
$H=p^2+4\ee^{-x}$] with energy $s_*^2$ starting from rest,
\beq
  \begin{split}
  x_{\rm cl}(t)&=2\log\frac{2\beta}{\pi}+2\log\cosh\frac{\pi t}{\beta},\\
  \ell_{\rm cl}(t)&=\ee^{x_{\rm cl}/2}=\frac{2\beta}{\pi}\cosh\frac{\pi t}{\beta},
  \end{split}
  \label{eq:xcl}
\eeq
the familiar growth of the renormalised length of the Einstein--Rosen
bridge~\cite{Harlow:2018tqv,Iliesiu:2021ari}; the reflection takes a time
$t\sim\beta/\pi$, after which the motion is ballistic,
$x_{\rm cl}\simeq2s_*t+2\log(\beta/\pi)$, with velocity $2s_*=2\pi/\beta$.

\subsection{The reflected packet and its freezing}
\label{sec:reflected}

\subsubsection{Bessel asymptotics and the outgoing wave}
\label{sec:bessel}

In the classically allowed region $x>x_{\rm turn}(s)=2\log(2/s)$, the
argument $z=4\ee^{-x/2}$ of the Bessel function is small compared to the
order, and $K_{2\ii s}$ becomes oscillatory. The precise statement, proved
in Appendix~\ref{app:bessel}, is that for $s>0$ fixed and $z\to0^+$
\beq
  \begin{split}
  K_{2\ii s}(z)={}&
  \sqrt{\frac{\pi}{2s\,\sinh(2\pi s)}}\;
  \cos\!\Big[\,2s\log\tfrac{2}{z}+\arg\Gamma(2\ii s)\,\Big]\\
  &\times\big(1+O(z^2)\big),
  \end{split}
  \label{eq:Kasym-general}
\eeq
which with $z=4\ee^{-x/2}$, i.e.\ $\log(2/z)=\tfrac x2-\log2$, reads
\beq
  \begin{split}
  K_{2\ii s}\!\big(4\ee^{-x/2}\big)&\simeq
  \sqrt{\frac{\pi}{2s\,\sinh(2\pi s)}}\;\cos\!\big[s\,x-\phi_0(s)\big],\\
  \phi_0(s)&=2s\log 2-\arg\Gamma(2\ii s).
  \end{split}
  \label{eq:Kasym}
\eeq
The reflection phase $\phi_0$ controls the outgoing wave. Its Stirling
expansion (Appendix~\ref{app:bessel}) is, for $s\gg1$,
\beq
  \begin{split}
  \arg\Gamma(2\ii s)&=2s\log(2s)-2s-\frac\pi4+O(s^{-1}),\\
  \phi_0(s)&=-2s\log s+2s+\frac\pi4+O(s^{-1}),
  \end{split}
  \label{eq:stirlingphase}
\eeq
so that
\beq
  \begin{split}
  \phi_0'(s)&=-2\log s+O(s^{-2}),\qquad \phi_0''(s)=-\frac2s+O(s^{-3}),\\
  \phi_0'''(s)&=\frac{2}{s^2}+O(s^{-4}).
  \end{split}
  \label{eq:phi0derivs}
\eeq
Inserting~\eqref{eq:Kasym} into~\eqref{eq:psi}, the measure factor
combines with the Bessel prefactor as
$2s\sinh(2\pi s)\sqrt{\pi/[2s\sinh(2\pi s)]}=\sqrt{2\pi s\sinh(2\pi s)}
\simeq\sqrt{\pi s}\,\ee^{\pi s}$ for $s\gg1$, so that, writing
$A(s)\equiv\sqrt{2\pi s\sinh(2\pi s)}\,\ee^{-\beta s^2/2}$, the time-evolved
state in the allowed region is
\beq
  \psi_t(x)=\mathcal N\!\int_0^\infty\!\dd s\;A(s)\,
  \ee^{-\ii t s^2}\,\cos\!\big[s x-\phi_0(s)\big],
  \label{eq:psiF}
\eeq
with $\mathcal N=\sqrt2/(\pi\sqrt{Z_\beta})$. The real envelope
$A(s)\simeq\sqrt{\pi s}\,\ee^{\pi s-\beta s^2/2}$ peaks at $s_*=\pi/\beta$
with amplitude width $1/\sqrt\beta$, and the entire effect of real-time
evolution is the chirp $\ee^{-\ii ts^2}$, a pure quadratic phase across
the energy band.

Split the cosine into its two exponentials, $\psi=\psi_++\psi_-$ with
$\psi_\pm\propto\int\dd s\,A(s)\,\ee^{-\ii ts^2}\ee^{\pm\ii[sx-\phi_0(s)]}$.
The stationary-phase condition for $\psi_\pm$ is $x=\phi_0'(s)\pm2st$,
i.e.\ $x=-2\log s\pm2st$ at large $s$. At $t=0$ both conditions place the
stationary point at $x=-2\log s_*=x_{\rm turn}(s_*)-2\log2$, inside the
wall, where~\eqref{eq:Kasym} does not hold: the decomposition into
counter-propagating waves is not available at $t=0$, in agreement with
Sec.~\ref{sec:HHpacket}. For $t>0$ the outgoing wave $\psi_+$ has its
stationary point at $x=2s_*t-2\log s_*$, which enters the allowed region
and thereafter coincides with the ballistic branch of~\eqref{eq:xcl}; the
incoming wave $\psi_-$ has its stationary point at $x=-2\log s_*-2s_*t$,
which recedes further into the wall, where the exact eigenfunctions
decay. Physically, $\psi_-$ is the
wave that has already reflected: for $t\gtrsim\beta$ only the outgoing
component survives in the allowed region,
\beq
  \psi_t(x)\simeq\frac{\mathcal N}{2}\int_0^\infty\!\dd s\;
  A(s)\,\ee^{\ii[\,sx-\phi_0(s)-ts^2\,]}\qquad(t\gg\beta).
  \label{eq:psiout}
\eeq
Expanding the phase about $s_*$ with~\eqref{eq:phi0derivs}, the cubic term
is $\tfrac16\phi_0'''(s_*)(s-s_*)^3=O(\beta^2)\times O(\beta^{-3/2})=O(\sqrt\beta)$
across the band, and the factor $\sqrt s$ in $A(s)$ varies by the same
relative amount. Thus~\eqref{eq:psiout} is, to leading order in $\beta$, a
Gaussian in $s$ with a quadratic phase --- a chirped Gaussian --- whose
Wigner function is again positive,
\beq
  \begin{split}
  W(x,p;t)\simeq\frac1\pi\exp\Big[&-\beta(p-s_*)^2\\
  &-\frac{\big(x-x_{\rm c}(t)-\tau(t)\,(p-s_*)\big)^2}{\beta}\Big],
  \end{split}
  \label{eq:Wout}
\eeq
with $x_{\rm c}(t)=2s_*t+2\log(\beta/\pi)$ the stationary point at $s_*$
and $\tau(t)=2t+\phi_0''(s_*)=2t-2\beta/\pi$ its dispersion,
$\dd x/\dd s|_{s_*}$. Equation~\eqref{eq:Wout} is a minimum-uncertainty
packet of momentum width $1/\sqrt{2\beta}$ [the width $\sigma_s$
of~\eqref{eq:saddle}], centred on the classical trajectory, and sheared
linearly in $t$; its phase-space area does not grow.

The same packet is obtained by inserting the two-wave split directly into
the Wigner integral~\eqref{eq:Wdef}, which also locates the cross term
that the split produces. For the $\psi_+\psi_+^*$ term, with
$\sigma=\tfrac12(s+s')$ and $\omega=s-s'$, the $y$ integral sets
$p=\sigma$; the amplitude product is
$A(s)A(s')\propto\ee^{-\beta(p-s_*)^2}\,\ee^{-\beta\omega^2/4}$, and the
phase is $\omega\,[x-\phi_0'(p)-2pt]+O(\omega^3\phi_0''')$. The
$\omega$ integral is then Gaussian and reproduces~\eqref{eq:Wout}: the
factor $\ee^{-\beta(p-s_*)^2}$ is the momentum profile, the
$\omega$-Gaussian gives the width $\sqrt{\beta/2}$ in $x$ at fixed $p$,
and the ridge $x=2pt+\phi_0'(p)=2pt-2\log p$ is the chirped centre line
$x_{\rm c}(t)+\tau(t)(p-s_*)$ to the order kept. For the
$\psi_+\psi_-^*$ cross term the $y$ integral instead sets
$p=\tfrac12(s-s')$, and the remaining phase
$2\sigma x-2\phi_0(\sigma)-4t\sigma p$ is stationary at
$x=\phi_0'(\sigma)+2pt\simeq-2\log s_*+2pt$: the cross term is the
interference pattern that $\psi_+$ would form with a left-moving partner,
centred at the turning point and at $|p|\lesssim1/\sqrt\beta$ for all
$t$, with an $x$-extent of order $t/\sqrt\beta$ comparable to that of the
packet. Since $\psi_-$ has no stationary point in the allowed region for
$t>0$, this pattern is not present in the exact state: it is an artefact
of continuing~\eqref{eq:Kasym} into the wall, and its support meets that
of~\eqref{eq:Wout} only in the Gaussian tails, at relative weight
$O(\ee^{-\pi^2/\beta})$. Equation~\eqref{eq:Wout} is therefore the complete
leading-order Wigner function of the reflected state, with no interference
ridge.

\subsubsection{Asymptotic freezing after the reflection}
\label{sec:freezing}

Any quantum evolution which starts from a Gaussian state and involves only
Clifford gates will keep the state within the restricted class of Gaussian
states. The shear in~\eqref{eq:Wout} is the phase-space image of this fact.
Let $H_0=\hat p^2$ be the free part of the Liouville
Hamiltonian~\eqref{eq:liouville}. Since $H_0$ is quadratic,~\eqref{eq:covariance}
applies with the classical flow $(x,p)\mapsto(x+2pt,p)$, which has unit
Jacobian, so for every state and every $t$
\beq
  \begin{split}
  W_{\ee^{-\ii H_0t}\psi}(x,p)&=W_\psi(x-2pt,\,p),\\
  \Neg\big[\ee^{-\ii H_0t}\psi\big]&=\Neg[\psi]\qquad\text{exactly}.
  \end{split}
  \label{eq:freeshear}
\eeq
Now let $H=\hat p^2+4\ee^{-\hat x}$ be the full Hamiltonian.
Because $V(x)=4\ee^{-x}$ decays exponentially in $x$ in the allowed region,
the M{\o}ller wave operator
$\Omega_+=\lim_{t\to\infty}\ee^{\ii Ht}\ee^{-\ii H_0t}$ exists on the
outgoing subspace~\cite{ReedSimon:1979} (Cook's criterion: free propagation
carries the packet to $x\sim2pt$ where $V\sim\ee^{-2pt}$ is integrable in
$t$), and for the reflected HH packet
\beq
  \begin{split}
  &\ee^{-\ii Ht}\,\psi_\beta
  \;=\;\ee^{-\ii H_0t}\,\psi_\beta^{\rm out}+\varepsilon(t),\\
  &\psi_\beta^{\rm out}\equiv\Omega_+^\dagger\psi_\beta,
  \qquad \|\varepsilon(t)\|\to0\quad(t\to\infty).
  \end{split}
  \label{eq:moller}
\eeq
Two caveats must be stated. First, $\Neg$ is not norm-continuous on
$L^2$, so~\eqref{eq:moller} alone does not bound the difference of
negativities. Second, the potential is nonzero at every finite $x$, so the
full evolution is never exactly quadratic at any finite time: the
exactness of~\eqref{eq:freeshear} does not transfer to $H$. The
statement we use is therefore semiclassical and asymptotic. In the WKB
regime the correction to free transport of $W$ is governed by the Moyal
tail of~\eqref{eq:moyaleq} with $h=V$, every term of which is
proportional to derivatives of $V$ on the support of $W$. By~\eqref{eq:Wout}
the packet sits at distance $x_{\rm c}(t)-x_{\rm turn}\simeq2s_*t$ from the
wall with width $\sigma_x(t)\simeq\sqrt{2/\beta}\,t$, so the weight of the
leading Gaussian within the wall region is $O(\ee^{-\pi^2/\beta})$ for all
$t\gg\beta$, uniformly in $t$. At leading semiclassical order
$\dd\Neg/\dd t$ is then negligible for $t\gg\beta/\pi$, and
\beq
  \Neg(t)\;\simeq\;\Neg\big[\psi_\beta^{\rm out}\big]
  =\text{const}
  \qquad(t\gg\beta/\pi),
  \label{eq:frozenvalue}
\eeq
at the disk level and at leading order.
At this order all change of the negativity occurs while the packet
interacts with the wall. Beyond leading order the relevant time is longer
than the reflection time $\beta/\pi$ of~\eqref{eq:xcl}: the band of the
HH state contains low-momentum components, of weight
$\ee^{-\beta(s_*-p)^2}$, which reach the wall late and reflect slowly, so
the subleading $\delta(\beta,t)$ is generated over the time during which
the slow tail of the band is still reflecting, and freezes thereafter.
The estimate $O(\ee^{-\pi^2/\beta})$ above concerns the $x$-tail of the
Gaussian packet; the slow components at $p=\alpha s_*$ carry the larger
weight $\ee^{-\pi^2(1-\alpha)^2/\beta}$, and it is this weight, not
$\ee^{-\pi^2/\beta}$, that sets the size of the frozen $\delta$
(Sec.~\ref{sec:numerics}).
The numerical evaluation of Sec.~\ref{sec:numerics} shows this directly:
$\delta$ is below $10^{-11}$ through the main reflection, rises over the
following few tens of $\beta/\pi$, and then approaches a frozen value,
roughly as $1/t$.
We do not control the remainder beyond leading order: a proof of constancy
in the limit $t\to\infty$ would require estimates of the Moyal series in
a weighted phase-space norm, including the Gaussian tails of $W$ and the
near-threshold components of the packet, and we do not attempt it.
Equation~\eqref{eq:frozenvalue} is to be read as the expected asymptotic
behaviour, supported by the leading-order calculation, not as an exact
finite-time constancy.

\subsection{The disk negativity}
\label{sec:diskN}

\subsubsection{\texorpdfstring{Time-reversal symmetry: no linear rise at $t=0$}{Time-reversal symmetry: no linear rise at t=0}}
\label{sec:parity}

The wavefunction~\eqref{eq:psi} is real, so
$\psi_{-t}=\psi_{t}^{\,*}$, and from the
definition~\eqref{eq:Wdef}
\beq
  W(x,p;-t)=W(x,-p;t)\qquad\Longrightarrow\qquad \Neg(-t)=\Neg(t).
  \label{eq:parity}
\eeq
The negativity of the HH state is an even function of $t$. It is also
differentiable at $t=0$: $\psi_\beta$ is analytic in the strip
$|\mathrm{Im}\,x|<\pi$ and decays at both ends, so $W(x,p;0)$ is
real-analytic and its zero set has measure zero, whence
$\dd\Neg/\dd t|_0=\int\mathrm{sgn}(W_0)\,\partial_tW|_0$. The
integrand is odd in $p$: $\mathrm{sgn}(W_0)$ is even in $p$
by~\eqref{eq:parity}, and the Moyal bracket~\eqref{eq:moyaleq} of two
functions even in $p$ --- here the symbol $h=p^2+4\ee^{-x}$ and $W_0$ ---
is odd in $p$, every term containing an odd number of $p$-derivatives.
Hence
\beq
  \Neg(t)=\Neg(0)+O(t^2)\qquad(t\to0):
  \label{eq:Nparity}
\eeq
there is no linear rise at $t=0$. Two separate things are established
here, evenness for all $t$ and differentiability at the origin; neither
by itself constrains the behaviour at later times, which is the subject of
Sec.~\ref{sec:freezing}. The universal law
$1+\tfrac{4b_1}{\pi}t$ derived for the discrete Wigner function
in~\cite{Basu:2024vem}, where the negativity always starts rising linearly
with a slope proportional to $b_1$, relies on the exact zeros of the
Wootters Wigner function~\cite{Wootters:1987} of a computational basis
state; it does not
transfer to the continuum, where the initial Wigner function
\eqref{eq:HHwigner} is strictly positive.

\subsubsection{The disk value}
\label{sec:diskvalue}

Combining~\eqref{eq:HHwigner}, \eqref{eq:Wout}, \eqref{eq:frozenvalue}
and~\eqref{eq:Nparity}, the genus-zero negativity of the HH state is
\beq
  \boxed{\;\Neg_\disc(\beta,t)=1+\delta(\beta,t),\qquad
  \delta(\beta,-t)=\delta(\beta,t),\;}
  \label{eq:Ndisk}
\eeq
with no linear term at $t=0$ by~\eqref{eq:Nparity}. Since $\psi_t$ is not
exactly Gaussian (its skewness is $O(\sqrt\beta)$, Appendix~\ref{app:packet}),
Hudson's theorem~\cite{Hudson:1974} implies $\delta(\beta,t)>0$ strictly
for every $t$; the question is its size. At leading semiclassical
order $\delta=0$ for all $t$: the Wigner function is the positive
Gaussian~\eqref{eq:HHwigner} at $t=0$ and the positive sheared
Gaussian~\eqref{eq:Wout} for $t\gg\beta/\pi$, and in between, at
leading order in the Moyal expansion of~\eqref{eq:moyaleq} with
$\hbar_{\rm eff}=\beta$, it is the Liouville transport of a positive
function. The reflection itself, $t\sim\beta/\pi$, is the part the
expansion controls least: the potential $4\ee^{-x}$ is not quadratic,
Wigner positivity is not preserved by the exact Moyal evolution, and the
Moyal corrections during the reflection are relatively $O(\sqrt\beta)$ in
the scaled variables, like the $O(\sqrt\beta)$ skewness of the initial
packet; both are polynomials in the scaled variables multiplying the
Gaussian, so at any fixed order in $\beta$ they can make $W$ negative only
in the Gaussian tails, where the fixed-order expansion is not uniform.
The expansion therefore predicts that $\delta$ is smaller than any power
of $\beta$, but does not fix its form; Sec.~\ref{sec:numerics} determines
it numerically.

A heuristic phase-space count makes the smallness natural. For a pure
state whose Wigner function is supported on a region of phase-space volume
$V$, Cauchy--Schwarz with the purity sum rule~\eqref{eq:moments} gives
\beq
  \Neg=\int|W|
  \;\le\;\Big(\!\int W^2\Big)^{1/2}\Big(\!\int_{\rm supp}\!1\Big)^{1/2}
  =\sqrt{\frac{V}{2\pi}} ;
  \label{eq:CS}
\eeq
the Gaussian packet has no compact support, so~\eqref{eq:CS} is not a
rigorous bound for it, but the sheared Gaussian~\eqref{eq:Wout} has an
effective area of one Wigner cell at all times, since the shear stretches
it without changing its area. In this heuristic sense the disk packet never
occupies more than $O(1)$ cells, which is why $\Neg_\disc=1+\delta$ with
$\delta$ small. A state is called classical if it has no negativity, and
semiclassical with respect to a given basis if its negativity does not
diverge as the Hilbert-space dimension grows~\cite{Balasubramanian:2026wkv}:
with respect to the length basis the HH state is, at genus zero,
classical up to the exponentially small $\delta$. $S_0$ does not enter
at genus zero, so whether this remains true at finite $\ee^{S_0}$ is not a
question the disk can answer (Sec.~\ref{sec:beyond}).

\subsubsection{Numerical evaluation of \texorpdfstring{$\delta$}{delta}}
\label{sec:numerics}

The problem is one-dimensional quantum mechanics, so $\delta$ can be
evaluated directly. We compute $\psi_\beta(x)$ from~\eqref{eq:psi-v} on
the line $\mathrm{Im}\,v=\pi/2$, where the factor $\ee^{2\pi^2/\beta}$
cancels analytically against $Z_\beta^{-1/2}$ and the integrand is $O(1)$,
so that the quadrature involves no large cancellations; the result
reproduces the exact norm and the exact mean energy~\eqref{eq:Emoments} to
$10^{-10}$. The state is then propagated with $H=\hat p^2+4\ee^{-\hat x}$
on a Fourier grid by a Chebyshev expansion of $\ee^{-\ii Ht}$, converged
to machine precision for the discretised Hamiltonian, and the Wigner
function~\eqref{eq:Wdef} is obtained by a fast Fourier transform in $y$.
Since free evolution is an exact shear [Eq.~\eqref{eq:freeshear}],
$\Neg(t)=\Neg[\ee^{\ii H_0t}\psi_t]$, and we evaluate the negativity on the
freely back-propagated state, which stays localised and is therefore
resolved at all times; at early times this agrees with the direct
evaluation on $\psi_t$. Halving the grid spacing changes $\delta$ by less
than $15\%$ wherever $\delta>10^{-10}$, an independent split-operator
propagation agrees to $5\%$, and the drift of the normalisation sets the
resolution at $\delta\simeq10^{-11}$.

\begin{figure*}[t]
\centering
\includegraphics[width=\textwidth]{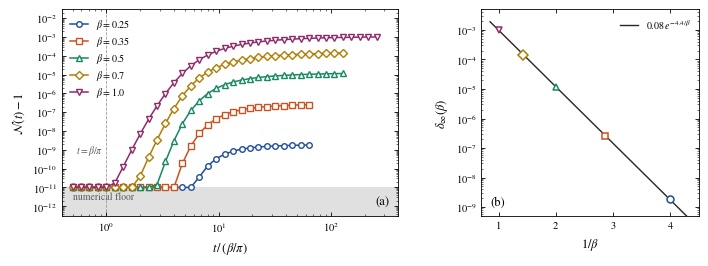}
\caption{Numerical genus-zero negativity of the HH state. (a)
$\Neg(t)-1=\delta(\beta,t)$ against $t$ in units of the reflection time
$\beta/\pi$, for $\beta=0.25$, $0.35$, $0.5$, $0.7$ and $1$; the shaded
band is the numerical floor and the dashed line the reflection time. (b)
The frozen value $\delta_\infty(\beta)$ against $1/\beta$, with the fit
$0.08\,\ee^{-4.4/\beta}$ (line).}
\label{fig:neg}
\end{figure*}

Figure~\ref{fig:neg} shows $\delta(\beta,t)$ for $\beta=0.25$--$1$.
Three features stand out. First, at $t=0$ and through the main
reflection -- up to $t\simeq\beta/\pi$ at $\beta=1$, $2\beta/\pi$ at
$\beta=0.5$ and $4\beta/\pi$ at $\beta=0.25$ -- $\delta$ is below
$10^{-11}$:
the negativity required by Hudson's theorem is far below the resolution
of a double-precision computation, even at $\beta=1$ where the corrections
to the Gaussian are not small. Second, $\delta$ then rises by many orders
of magnitude over the following few tens of $\beta/\pi$ and approaches a
final value, roughly as $1/t$, reaching it to within $5\%$ at the longest
time computed ($256\,\beta/\pi$ at $\beta=1$). The negative part of the back-propagated Wigner
function is localised within a few units of the turning point, at
momenta $|p|\simeq s_*/2$, many widths $1/\sqrt{2\beta}$ below the
carrier and nowhere near the outgoing packet: it is produced where the
low-momentum tail of the band meets the wall, long after the centre of the
packet has left, as anticipated in Sec.~\ref{sec:freezing}. Third, the
frozen values, extrapolated from the last two times assuming a $1/t$
approach,
\beq
  \begin{split}
  \delta_\infty(0.25)&=2.2\times10^{-9},\quad \delta_\infty(0.35)=2.9\times10^{-7},\\
  \delta_\infty(0.5)&=1.3\times10^{-5},\quad\ \ \delta_\infty(0.7)=1.6\times10^{-4},\\
  \delta_\infty(1.0)&=1.1\times10^{-3},
  \end{split}
  \label{eq:deltainf}
\eeq
are reproduced to within $7\%$ by
$\delta_\infty\simeq0.08\,\ee^{-4.4/\beta}$: an exponential in $1/\beta$,
not a power or a stretched exponential. This is what the tail mechanism
suggests: the band carries weight
$\ee^{-\beta(s_*-p)^2}=\ee^{-\pi^2(1-\alpha)^2/\beta}$ at momentum
$p=\alpha s_*$, exponential in $1/\beta$ at fixed $\alpha$, and
$c\simeq4.4$ corresponds to $\alpha\simeq0.34$, of the same order as
the position $|p|\simeq s_*/2$ of the negative lobes. These numbers replace the expectation
of Sec.~\ref{sec:diskvalue} by a measurement: at $\beta=0.25$ the HH
state is Wigner-positive to one part in $10^{9}$ at all times.

\subsection{Remarks on the double trumpet}
\label{sec:beyond}
\label{sec:noramp}

The connected double trumpet produces a ramp in the SFF; does it affect
$\Neg(t)$? Two remarks are all the genus-zero analysis supports, and we
state them together with the reason it supports nothing more.

\paragraph{The mean Wigner function.} Expanding the density matrix in
energy, with $\chi_E$ the unnormalised Bessel function~\eqref{eq:chidef}
and $\rho_0(E)=\sinh(2\pi\sqrt E)$, Eq.~\eqref{eq:psi} gives
\beq
  \begin{split}
  \rho_\psi(x,x';t)=\frac{2}{\pi^2Z_\beta}\!\int\!\dd E\,\dd E'\,&\rho_0(E)\rho_0(E')\,
  \chi_E(x)\chi_{E'}(x')\\
  &\times\ee^{-(\beta/2+\ii t)E}\,\ee^{-(\beta/2-\ii t)E'},
  \end{split}
  \label{eq:rhopsi}
\eeq
whose trace is one by the orthogonality relation of
Appendix~\ref{app:norm}. In the matrix-model completion the product
$\rho_0(E)\rho_0(E')$ is replaced by the averaged two-level density, whose
connected part at genus zero is the double trumpet~\cite{Saad:2019lba},
$Z_\conn(\beta_1,\beta_2)=\sqrt{\beta_1\beta_2}/[2\pi(\beta_1+\beta_2)]$ in
Laplace space; continued to $\beta_{1,2}=\beta\pm\ii t$,
\beq
  Z_\conn(\beta+\ii t,\beta-\ii t)=\frac{\sqrt{\beta^2+t^2}}{4\pi\beta}
  \;\simeq\;\frac{t}{4\pi\beta}\qquad(t\gg\beta),
  \label{eq:ramp}
\eeq
the linear ramp. Relative to the disconnected kernel of~\eqref{eq:rhopsi}
the connected kernel is therefore suppressed by
\beq
  \frac{Z_\conn(\beta+\ii t,\beta-\ii t)}{Z^{\rm JT}(\beta)^2}
  \;\sim\;t\,\ee^{-2S_0}\qquad(t\gg\beta),
  \label{eq:connsize}
\eeq
by~\eqref{eq:ZJT}. For the SFF this ratio is what matters, because the
disconnected term, the exact survival probability~\eqref{eq:survival2},
decays as $(\beta/t)^3\ee^{-2\pi^2/\beta}$: the ramp overtakes it at
\beq
  t_{\rm dip}=\beta^{1/4}\,\ee^{S_0/2}
  \label{eq:tdip}
\eeq
(for the SFF normalised by the disk partition function, in the
double-trumpet approximation), and is exposed from there to the
Heisenberg time $t_{\rm H}\sim\ee^{S_0}$~\cite{Saad:2019lba,Okuyama:2019xbv}.
For the averaged Wigner function the disconnected term is the disk packet
of Sec.~\ref{sec:reflected}, positive, of unit weight and frozen at
leading order, which does not decay; so the double trumpet remains a
relative $O(t\,\ee^{-2S_0})$ correction to $\avg{W}$ near the packet for
all $t\ll\ee^{S_0}$. This is a weak statement -- an $O(1)$ term dominating
an $\ee^{-2S_0}$ one -- and it is pointwise, not an $L^1$ bound: the
$1/(E_1-E_2)^2$ behaviour of the connected density becomes, in the
variable conjugate to $E_1-E_2$, a growth linear in the distance from the
classical ridge (the phase-space image of the ramp), so the connected
correction to the kernel is not integrable over $x$.

\paragraph{Why nothing follows for the negativity.} Indeed, the absolute
value creates trouble -- while calculating the ensemble average of the
Wigner function and its powers is straightforward using the gravity path
integral, the absolute value is a non-analytic function, and $\avg{|W|}$ is
controlled by the fluctuations of $W$ rather than by its
mean~\cite{Basu:2025kjr,Bhattacharya:2026bda}. More basically, the
observable itself is not defined beyond genus zero. At finite $\ee^{S_0}$
the spectrum is discrete, the length-basis states are overcomplete, with
null states~\cite{Iliesiu:2024cnh}, and the wavefunction
$\sum_nc_n\chi_{E_n}(x)$ of any state is an almost-periodic sum of
standing waves at large $x$: it is not square integrable in $x$, and
$\int|W|\,\dd x\,\dd p$ diverges. A negativity at finite $\ee^{S_0}$
therefore requires a different computational basis, such as the Krylov
chain generated by the HH seed (Sec.~\ref{sec:spread}), which is the
non-perturbative version of the length basis~\cite{Basu:2025kjr} and is a
genuine basis at any $\ee^{S_0}$. Any departure from the genus-zero value
$1+\delta$ is in that sense non-perturbative in the topological
expansion, and we make no statement about it.

\section{Spread complexity on the same footing}
\label{sec:spread}

Given the input $(H,\ket{\psi_\beta})$, we can construct a basis for the
Hilbert space -- more precisely, for the subspace spanned by states of the
form $\ee^{-\ii tH}\psi_\beta$ -- called the Krylov
basis~\cite{Parker:2018yvk,Balasubramanian:2022tpr}
(see~\cite{Rabinovici:2020ryf,Caputa:2021sib,Balasubramanian:2022dnj} and
the reviews~\cite{Nandy:2024htc,Baiguera:2025dkc}). In other words, we
consider the set of states $\psi_\beta,H\psi_\beta,H^2\psi_\beta\cdots$ and
build an orthonormal basis from it using the Gram-Schmidt procedure. The
resulting basis is called the Krylov basis. Note that the Krylov basis is an
ordered basis; we will henceforth use the notation $\ket{K_n}$ to denote the
$n$-th Krylov state, with $\ket{K_0}=\ket{\psi_\beta}$. One of the key
features of the Krylov basis is that the Hamiltonian is tri-diagonalized in
this basis, i.e., $H\ket{K_n}=b_n\ket{K_{n-1}}+a_n\ket{K_n}+b_{n+1}\ket{K_{n+1}}$.
These numbers are called the Lanczos or Krylov coefficients.
The Krylov basis has the nice property that it minimizes the ``spreading''
of the wavefunction under time evolution~\cite{Balasubramanian:2022tpr}.
This can be quantified as follows: consider the time evolved state
$\ket{\psi(t)}=\ee^{-\ii tH}\ket{\psi_\beta}$. For any ordered, orthonormal
basis $B=\{\ket{n_B}\}$, we can define a measure of the spread of the state
$\psi(t)$, $\Cplx_B(t)=\sum_n c_n|\langle n_B|\psi(t)\rangle|^2$, where
$c_n$ is any positive, monotonically increasing weight function; for e.g., a
natural choice is to take $c_n=n$. We may think of $\Cplx_B(t)$ as some
notion of ``spread complexity'', i.e., a measure of how spread out the state
is in the given choice of basis. In the Krylov basis,
\beq
  \Cplx(t)\;=\;\sum_{n\ge0}n\,\big|\langle K_n|\psi(t)\rangle\big|^2 .
  \label{eq:Cdef}
\eeq
For the HH state the relevant spectral measure is the thermal one,
$p(E)\,\dd E=\rho(E)\ee^{-\beta E}\dd E/Z^{\rm JT}(\beta)$, and all Krylov
data are functionals of $p$; the time $t$ is the boundary time
of~\eqref{eq:psit}, as everywhere in this paper. Spread complexity of
thermal and thermofield-double states in JT gravity, the Schwarzian
theory and related models has been studied
in~\cite{Rabinovici:2023yex,Erdmenger:2023wjg,Heller:2024ldz} (see
also~\cite{Jian:2020qpp} for operator growth in JT); the survival
amplitude and the Laguerre chain below are standard, and what is new here
is the exact $t^4$ coefficient~\eqref{eq:C4} and the subleading Lanczos
law~\eqref{eq:chainsub}.

\subsection{Exact survival amplitude and exact early growth}
\label{sec:survival}

The integral $Z_\gamma=\int_0^\infty2s\sinh(2\pi s)\ee^{-\gamma s^2}\dd s$
converges absolutely and locally uniformly for $\mathrm{Re}\,\gamma>0$ and
defines a holomorphic function there; the closed form
$\pi^{3/2}\gamma^{-3/2}\ee^{\pi^2/\gamma}$ (principal branch) is likewise
holomorphic and agrees with the integral on the positive real axis
by~\eqref{eq:Zdef}. By the identity theorem the two coincide on the half
plane, and setting $\gamma=\beta+\ii t$ gives the exact survival amplitude
\beq
  \begin{split}
  S(t)&\;\equiv\;\langle\psi_\beta|\ee^{-\ii Ht}|\psi_\beta\rangle
  =\int_0^\infty\!\dd E\,p(E)\,\ee^{-\ii Et}
  =\frac{Z_{\beta+\ii t}}{Z_\beta}\\
  &=\Big(\frac{\beta}{\beta+\ii t}\Big)^{3/2}
   \exp\Big[\frac{\pi^2}{\beta+\ii t}-\frac{\pi^2}{\beta}\Big].
  \end{split}
  \label{eq:survival}
\eeq
Since
$\mathrm{Re}\big[\pi^2/(\beta+\ii t)-\pi^2/\beta\big]
=-\pi^2t^2/[\beta(\beta^2+t^2)]$,
\beq
  \begin{split}
  |S(t)|^2&=\Big(\frac{\beta^2}{\beta^2+t^2}\Big)^{3/2}
  \exp\Big[-\frac{2\pi^2t^2}{\beta(\beta^2+t^2)}\Big]\\
  &=1-\sigma_E^2\,t^2+O(t^4),
  \end{split}
  \label{eq:survival2}
\eeq
with $\sigma_E^2=2\pi^2/\beta^3+3/(2\beta^2)$ the exact thermal energy
variance of~\eqref{eq:Emoments}.

The first Lanczos coefficients are fixed by the moments $\mu_k$ of $H$ in
the seed: $a_0=\langle K_0|H|K_0\rangle=\mu_1$ and
$b_1^2=\|(H-a_0)K_0\|^2=\mu_2-\mu_1^2$, so that
\beq
  \begin{split}
  a_0&=\langle E\rangle_\beta=\frac{\pi^2}{\beta^2}+\frac{3}{2\beta},\\
  b_1&=\sigma_E=\sqrt{\frac{2\pi^2}{\beta^3}+\frac{3}{2\beta^2}}
  =\frac{\sqrt2\,\pi}{\beta^{3/2}}\Big(1+O(\beta)\Big).
  \end{split}
  \label{eq:ab-exact}
\eeq
The early growth of the spread complexity is exactly quadratic: the
amplitudes $\varphi_n(t)=\langle K_n|\ee^{-\ii Ht}|K_0\rangle$ satisfy
$\varphi_n(t)=O(t^n)$, since each application of $H$ moves at most one step
up the chain, and
$\varphi_1(t)=-\ii b_1t-\tfrac{t^2}{2}b_1(a_0+a_1)+O(t^3)$, so the
$O(t^3)$ term of $|\varphi_1|^2$ is the interference of $-\ii b_1t$ with a
real $t^2$ term and cancels, while sites $n\ge2$ contribute $O(t^4)$:
\beq
  \Cplx(t)\;=\;\sigma_E^2\,t^2\;+\;O(t^4).
  \label{eq:earlyC}
\eeq
The $O(t^4)$ term is also exact. Carrying the expansion one order further
(Appendix~\ref{app:cumulants}), with $\kappa_r$ the cumulants of the
thermal energy distribution, $\kappa_r=(-1)^r\partial_\beta^r\log Z_\beta$,
the first three Lanczos data are $b_1^2=\kappa_2$, $a_1-a_0=\kappa_3/\kappa_2$
and $b_2^2=\kappa_4/\kappa_2+2\kappa_2-\kappa_3^2/\kappa_2^2$, and
\beq
  \Cplx(t)=\kappa_2t^2+\Big(\frac{\kappa_4}{6}-\frac{\kappa_3^2}{4\kappa_2}\Big)t^4+O(t^6).
  \label{eq:C4general}
\eeq
For the JT measure, $\log Z_\beta=\text{const}-\tfrac32\log\beta+\pi^2/\beta$
gives $\kappa_2=(4\pi^2+3\beta)/2\beta^3$, $\kappa_3=3(2\pi^2+\beta)/\beta^4$,
$\kappa_4=3(8\pi^2+3\beta)/\beta^5$, and hence
\beq
  \Cplx(t)=\Big(\frac{2\pi^2}{\beta^3}+\frac{3}{2\beta^2}\Big)t^2
  -\frac{2\pi^4}{\beta^5(4\pi^2+3\beta)}\,t^4+O(t^6).
  \label{eq:C4}
\eeq
The negative $t^4$ coefficient shows that the growth is slower than the
pure quadratic law from the start; for the Laguerre reference measure of
Sec.~\ref{sec:laguerre} the same combination $\kappa_4/6-\kappa_3^2/4\kappa_2$
vanishes identically, consistently with its exactly quadratic spread.
The contrast with the negativity is this: the spread complexity grows
quadratically with the exact coefficient $\sigma_E^2$, whereas the
negativity has no linear term by~\eqref{eq:Nparity}, and its $O(t^2)$
coefficient, $\tfrac12\delta''(\beta,0)$, vanishes at leading
semiclassical order and is otherwise not computed; we expect it to be
smaller than any power of $\beta$ because negativity first appears in the
Gaussian tails, but this is an expectation, not a consequence of
Eqs.~\eqref{eq:Nparity} and~\eqref{eq:Ndisk}.

\subsection{An exactly solvable reference chain and the Schwarzian asymptotics}
\label{sec:laguerre}

Near its edge the JT density behaves as
$\rho\propto\sinh(2\pi\sqrt E)\simeq 2\pi\sqrt E$, so the weight
$E^{1/2}\ee^{-\beta E}$ is the natural edge-truncated reference, and its
chain is exactly solvable (Appendix~\ref{app:laguerre}): for
$p_L(E)\propto E^{\alpha}\ee^{-\beta E}$ on $[0,\infty)$ (the JT edge is
$\alpha=\tfrac12$),
\beq
  \begin{split}
  &a_n=\frac{2n+\alpha+1}{\beta},\qquad
  b_n=\frac{\sqrt{n(n+\alpha)}}{\beta},\\
  &\Cplx_L(t)=(\alpha+1)\,\frac{t^2}{\beta^2}=b_1^2\,t^2
  \quad\text{for \emph{all} }t.
  \end{split}
  \label{eq:laguerre}
\eeq
The Lanczos coefficients of the full JT measure
$p(E)\propto\sinh(2\pi\sqrt E)\,\ee^{-\beta E}$ approach the Laguerre law
at large $n$,
\beq
  a_n=\frac{2n}{\beta}\big(1+o(1)\big),\qquad
  b_n=\frac{n}{\beta}\big(1+o(1)\big),
  \label{eq:chainasym}
\eeq
with the crossover away from the Schwarzian-peak values~\eqref{eq:ab-exact}
around $n_*\sim\beta\langle E\rangle_\beta/2\simeq\pi^2/2\beta$. The
argument: writing the weight as $\ee^{-Q(E)}$ with
$Q(E)=\beta E-2\pi\sqrt E+O(\ee^{-4\pi\sqrt E})$, the large-$n$ recurrence
coefficients for exponential weights on $[0,\infty)$ are governed by the
Mhaskar--Rakhmanov--Saff numbers of $Q$; for the linear part $Q_0=\beta E$
these give exactly~\eqref{eq:laguerre}, while the subexponential correction
$-2\pi\sqrt E$ shifts the equilibrium endpoints by $O(\sqrt n)$ relative to
their $O(n)$ size, leaving~\eqref{eq:chainasym}
unchanged~\cite{SaffTotik:1997}. Explicitly, for a weight $\ee^{-Q(E)}$
on $[0,\infty)$ the right endpoint $B_n$ of the equilibrium support of the
degree-$n$ problem is fixed by~\cite{SaffTotik:1997,LevinLubinsky:2001}
(here $\ee^{-Q}$ is the orthogonality weight itself; in the convention
of~\cite{LevinLubinsky:2001}, where the weight is $W^2=\ee^{-2Q}$, the
prefactor is $1/\pi$)
\beq
  n=\frac{1}{2\pi}\int_0^{B_n}\!\dd E\;Q'(E)\sqrt{\frac{E}{B_n-E}}\,,
  \label{eq:MRS}
\eeq
which for $Q=\beta E-2\pi\sqrt E$ gives $n=\beta B_n/4-\sqrt{B_n}$ (the
two terms integrate to $\beta B_n/4$ and $-\sqrt{B_n}$); with
$a_n\simeq B_n/2$ and $b_n\simeq B_n/4$ for a one-sided support this gives
the subleading law
\beq
  a_n=\frac{2n}{\beta}+\frac{4\sqrt n}{\beta^{3/2}}+O(1),
  \qquad
  b_n=\frac{n}{\beta}+\frac{2\sqrt n}{\beta^{3/2}}+O(1),
  \label{eq:chainsub}
\eeq

\subsection{Disk-level growth: what is and is not established}
\label{sec:spread-summary}

Writing the disk-level spread complexity as
\beq
  \Cplx(t)\;=\;c(t)\,\frac{t^2}{\beta^2},
  \qquad
  c(0)=\big(\beta\sigma_E\big)^2=\frac{2\pi^2}{\beta}+\frac32,
  \label{eq:Cglobal}
\eeq
the exact short-time law~\eqref{eq:C4} gives
$c(t)=c(0)-2\pi^4t^2/[\beta^3(4\pi^2+3\beta)]+O(t^4)$, a coefficient that
decreases from its Schwarzian-peak value. What happens at later times is
not derived in this paper. The leading Lanczos asymptotics~\eqref{eq:chainasym}
coincide with those of the Laguerre chain~\eqref{eq:laguerre}, whose
growth is exactly quadratic for all $t$, and this suggests that $c(t)$
remains of order one; but the subleading $O(\sqrt n)$ terms
of~\eqref{eq:chainsub} dominate the $O(1)$ Laguerre offset $\alpha/2$, and
Eq.~\eqref{eq:C4} already shows that the reference chain cannot be
substituted for the full measure. A derivation of the late-time scaling
of $\Cplx$, or a converged evolution on the actual recurrence
coefficients, is required before the growth can be called quadratic
throughout, and we make no such claim. We note only the structural
contrast with the bounded spectra of double-scaled models, where Krylov
complexity measures the wormhole
length~\cite{Rabinovici:2023yex,Lin:2022rbf,Xu:2024gfm} and the chain
coefficients saturate, whereas the unbounded genus-zero Schwarzian
spectrum makes $b_n\simeq n/\beta$ grow indefinitely.

At genus zero the wavefunction spreads without bound along the chain,
exactly as the geodesic length does, while the negativity of the Wigner
function stays at $1+o(1)$ [Eq.~\eqref{eq:Ndisk}]: the spreading of the
state over the Krylov basis is, at the disk level, driven by essentially
classical dynamics, and large spread complexity does not require
negativity~\cite{Balasubramanian:2026wkv}. Beyond the disk, the number of
Krylov sites is set by the rank of the spectral support of the seed, so at
finite $\ee^{S_0}$ the chain terminates and the spread complexity must
eventually saturate~\cite{Balasubramanian:2022tpr}; see
also~\cite{Balasubramanian:2024lqk,Iliesiu:2021ari,Stanford:2022fdt} for the
analogous saturation of the wormhole size itself. Neither the saturation value nor
the time at which it is reached is addressed in this paper.

\subsection{Two scales: carrier momentum versus band width}
\label{sec:twoscales}

The continuum HH state carries two distinct scales, which coincide in the
discrete $\beta=0$ model but bifurcate at finite temperature. The first is
the carrier momentum $s_*=\pi/\beta$, the mean of the momentum
distribution of the reflected packet~\eqref{eq:Wout}, which sets the
velocity of the Einstein--Rosen bridge~\eqref{eq:xcl} and the reflection
time $t\sim\beta/\pi$ during which the disk corrections $\delta$ are
generated. The
second is the band width $\sigma_E\sim\sqrt2\pi/\beta^{3/2}$, the first
Lanczos coefficient of the spread chain [Eq.~\eqref{eq:ab-exact}], which
sets the survival decay and the early spread growth. In the discrete model
of~\cite{Basu:2024vem,Basu:2025kjr} at infinite temperature, in the
large-$D$ limit, the average values of the Krylov coefficients are given by
$a_n=0$, $b_n=1$ ($n$ fixed, $D\to\infty$), with standard deviations of
$O(1/\sqrt D)$: there the carrier, the band width, and $b_1$ are all
$O(1)$ and indistinguishable. In the continuum the two roles separate
cleanly: the carrier fixes how fast the bridge grows, and the band width
fixes how fast the state leaves its seed.

\section{Seed-normalised negativity on the same footing}
\label{sec:normneg}
\label{sec:seednorm}

A natural normalisation of the negativity, following~\cite{Basu:2026inf},
divides out the decaying return amplitude. Let
$\ket{\widetilde\psi(t)}\equiv\ket{\psi(t)}/S(t)$ be the seed-normalised
vector (unit amplitude on the seed $\ket{K_0}=\ket{\psi_\beta}$ for all
$t$, hence not a normalised state), with $S(t)$ the survival
amplitude~\eqref{eq:survival}. The Wigner transform is quadratic in the
vector, so $\psi\to\psi/S$ gives $W\to W/|S|^2$ pointwise and
\beq
  \Nnorm(\beta,t)\;=\;\frac{\Neg(\beta,t)}{\Psurv(\beta,t)}\,,
  \qquad
  \Psurv(\beta,t)\equiv|S(t)|^2 ,
  \label{eq:seednorm}
\eeq
exactly, in any frame and at any genus: $\Nnorm$ is a derived diagnostic,
the Wigner $L^1$ norm of an unnormalised vector, and does not inherit the
operational meaning of $\Neg$. The rate $\NegRate(t)\equiv\dd\log\Nnorm/\dd t$
of~\cite{Basu:2026inf} splits exactly as
$\NegRate=\dd\log\Neg/\dd t-\dd\log\Psurv/\dd t$, and with the exact
survival~\eqref{eq:survival2}, for all $t$,
\beq
  \begin{split}
  -\frac{\dd\log\Psurv}{\dd t}
  &=\frac{t\,\big(3\beta^2+4\pi^2\beta+3t^2\big)}{(\beta^2+t^2)^2}\\
  &=
  \begin{cases}
    2\sigma_E^2\,t\,\big(1+O(t^2/\beta^2)\big), & t\ll\beta,\\[2mm]
    \dfrac{3}{t}+\dfrac{4\pi^2\beta-3\beta^2}{t^3}+O(t^{-5}), & t\gg\sqrt\beta,
  \end{cases}
  \end{split}
  \label{eq:rate}
\eeq
with $\sigma_E^2$ the exact variance of~\eqref{eq:Emoments}; the pure
$3/t$ law requires $t\gg\sqrt\beta$, not merely $t\gg\beta$. At genus
zero the numerator of~\eqref{eq:seednorm} is $1+\delta$
by~\eqref{eq:Ndisk}, so
\beq
  \boxed{\;
  \begin{aligned}
  \Nnorm_\disc(\beta,t)&=\Big(1+\frac{t^2}{\beta^2}\Big)^{3/2}
  \exp\!\Big[\frac{2\pi^2t^2}{\beta(\beta^2+t^2)}\Big]\\
  &\quad\times\big(1+\delta(\beta,t)\big),
  \end{aligned}\;}
  \label{eq:Nnormdisk}
\eeq
with the two limits
\beq
  \begin{split}
  &\Nnorm_\disc(\beta,t)=1+\sigma_E^2\,t^2+O(\sigma_E^4t^4)\qquad(\sigma_Et\ll1),\\
  &\Nnorm_\disc(\beta,t)\simeq\frac{t^3}{\beta^3}\,\ee^{2\pi^2/\beta}\,
  \big(1+O(\beta/t^2)\big)\\
  &\qquad\text{and}\quad
  \NegRate(t)\simeq\frac{3}{t}\qquad(t\gg\sqrt\beta),
  \end{split}
  \label{eq:cubic}
\eeq
where $\delta$ enters only at the level of Sec.~\ref{sec:numerics}. Both
laws are properties of the inverse survival probability, not newly
generated negativity. The late law reads off the spectral edge: for any
measure with $p(E)\propto E^{\alpha}$ as $E\to0$, Watson's lemma gives
$\Psurv\sim t^{-2(\alpha+1)}$ and $\NegRate\to2(\alpha+1)/t$; the JT edge
$\rho\propto\sqrt E$ ($\alpha=\tfrac12$) gives the $3/t$
of~\eqref{eq:cubic}, with no oscillatory corrections. This is the JT
analogue of the dimension-reading property of the normalised rate in
AdS$_3$~\cite{Basu:2026inf}, with the conformal weight replaced by the
edge exponent. Beyond genus zero the denominator carries the double
trumpet, whose ramp overtakes the disk term at $t_{\rm dip}$
of~\eqref{eq:tdip}, so Eq.~\eqref{eq:Nnormdisk} is to be used only up to
$t_{\rm dip}$; the numerator is not defined there (Sec.~\ref{sec:beyond}),
and we make no statement beyond it.

\section{Summary of results, with their status}
\label{sec:master}

We collect the results of the paper, labelling each by its status:
\emph{exact} (an identity), \emph{leading} (a leading statement in the
semiclassical parameter $\beta$) or \emph{numerical}
(Sec.~\ref{sec:numerics}). For the negativity of the Hartle--Hawking state
at genus zero,
\beq
  \begin{split}
  \Neg(\beta,t)&=1+\delta(\beta,t),\qquad
  \delta(\beta,-t)=\delta(\beta,t),\\
  \delta(\beta,t)&<10^{-11}\ \ (\text{main reflection}),\\
  \delta(\beta,t)&\to\delta_\infty(\beta)\simeq0.08\,\ee^{-4.4/\beta},
  \end{split}
  \label{eq:master-all}
\eeq
the first line exact and the second numerical for $0.25\le\beta\le1$,
with the frozen value approached over a few tens of $\beta/\pi$ and
reached to within $5\%$ at $256\,\beta/\pi$ (at $\beta=1$). The only expansion parameter is
$\beta$; $S_0$ does not enter. There is no linear rise at $t=0$
[Eq.~\eqref{eq:Nparity}, exact], and the double trumpet is a relative
$O(t\,\ee^{-2S_0})$ correction to the mean Wigner function
(Sec.~\ref{sec:beyond}); beyond genus zero the negativity is not defined
in the length basis, and nothing is claimed. The time scales fixed by the
genus-zero data are the reflection time $\beta/\pi$ of~\eqref{eq:xcl},
the freezing time of $\delta$, and, for the SFF, the dip
time~\eqref{eq:tdip}. For $\beta\gtrsim1$ ($s_*\lesssim1$) the
semiclassical machinery fails, the HH state being dominated by
near-threshold energies in the Airy regime of the Bessel functions, and
the low-temperature case is left open. Table~\ref{tab:compare} collects
the four probes.

\begin{table*}[t]
\centering
\renewcommand{\arraystretch}{1.6}
\begin{tabular}{lp{3.6cm}p{7.0cm}p{4.4cm}}
\hline\hline
Observable & Early law & Later behaviour at genus zero & Status \tabularnewline
\hline
$\Neg$ & \raggedright $1+\delta$, $\delta<10^{-11}$, even in $t$ &
  \raggedright rises over a few tens of $\beta/\pi$, then approaches $\delta_\infty\simeq0.08\,\ee^{-4.4/\beta}$ &
  \raggedright evenness exact; values numerical \tabularnewline
$\Nnorm=\Neg/\Psurv$ & \raggedright $1+\sigma_E^2t^2$ &
  \raggedright $(1+\delta)/\Psurv_\disc\simeq(t/\beta)^3\ee^{2\pi^2/\beta}$, $\NegRate\simeq3/t$; valid to $t_{\rm dip}$ &
  \raggedright exact given $\delta$ \tabularnewline
$\Cplx$ & \raggedright $\sigma_E^2t^2-\dfrac{2\pi^4t^4}{\beta^5(4\pi^2+3\beta)}$ &
  \raggedright $c(t)\,t^2/\beta^2$ with $c(t)$ not derived; $a_n,b_n$ of Eq.~\eqref{eq:chainsub} &
  \raggedright $t^4$ law exact; $a_n,b_n$ to leading order \tabularnewline
SFF $=\overline{\Psurv}$ & \raggedright $1-\sigma_E^2t^2$ &
  \raggedright $(\beta/t)^3\ee^{-2\pi^2/\beta}$, then the ramp~\eqref{eq:ramp} from $t_{\rm dip}=\beta^{1/4}\ee^{S_0/2}$ &
  \raggedright exact; ramp from the double trumpet \tabularnewline
\hline\hline
\end{tabular}
\caption{The probes of the HH state considered in this paper, with the
status of each entry. Only the SFF has a controlled description beyond
genus zero (its double-trumpet ramp); for the negativity nothing beyond
genus zero is claimed.}
\label{tab:compare}
\end{table*}

\section{Discussion}
\label{sec:disc}

\paragraph{Summary.} In this work, we have studied the growth of Wigner
negativity under time evolution of the Hartle--Hawking state in JT gravity.
At genus zero and to leading order in the semiclassical parameter $\beta$
the HH state is a Gaussian wavepacket at rest at the turning point of the
Liouville wall, and its evolution is a reflection followed by a Clifford
shear; its negativity is therefore $1+\delta$ with $\delta$ beyond all
orders in $\beta$, even in $t$ and with no linear term at $t=0$. A direct
numerical evaluation finds $\delta<10^{-11}$ through the reflection and a
frozen late value $\delta_\infty\simeq0.08\,\ee^{-4.4/\beta}$, generated by
the low-momentum tail of the packet. The exact survival amplitude fixes
the spread complexity to order $t^4$ and the seed-normalised diagnostic
throughout the genus-zero era. The double trumpet corrects the mean Wigner
function only at relative order $t\,\ee^{-2S_0}$, but the length-basis
negativity itself is not defined beyond genus zero, and we claim nothing
about finite $\ee^{S_0}$. We end with some remarks on the outlook and
future directions.

\paragraph{Consistency with the Krylov--Wigner programme.} The numerical
results of~\cite{Basu:2024vem} are consistent with the claim that the
Krylov basis minimizes the early-time growth of Wigner negativity, and in
the Krylov basis, the Wigner negativity grows slowly, as a power law for a
long time, and becomes exponentially large only at exponentially long
times. The continuum length basis of JT sharpens the Krylov-basis
statement at genus zero: the negativity does not grow at all beyond the
exponentially small $\delta$, Eq.~\eqref{eq:Ndisk}. A crucial point is
that the Wigner function and its negativity is defined with respect to a
choice of computational basis: the two statements concern different
objects, the discrete Wootters negativity in the chain basis and the
continuous-variable negativity in $(x,p)$, and the negativity is invariant
under metaplectic maps~\eqref{eq:covariance} but not under the nonlinear,
discretising change of frame between them. Whether the bulk-frame
negativity grows as a power law beyond the genus-zero era, as the Krylov
frame does, is not fixed by either argument.

\paragraph{Probes and clocks.} In recent work, the Shannon entropy and the
first moment of the above distributions have been studied as measures of
the spreading of a quantum state under time evolution; these measures are
commonly called spread complexity or Krylov
complexity~\cite{Balasubramanian:2022tpr}. Here all the probes are built
from the same state and the same spectral data, yet they respond to the
spectrum on different time scales: the negativity freezes within a few
tens of $\beta/\pi$, the survival probability decays until its dip at
$t_{\rm dip}=\beta^{1/4}\ee^{S_0/2}$, and the spread complexity grows
without bound at genus zero.

\paragraph{What would ramp.} It would also be interesting to consider other
initial states, such as for example the thermofield double state, and
observables whose disconnected term decays in time, as the survival
probability does. For the SFF it is the decay of the disk term, not the
time labels of the boundaries, that exposes the double trumpet. Whether
the same holds for other functionals of the state is a case-by-case
question that requires the corresponding connected correlators; the
seed-normalised negativity of Sec.~\ref{sec:normneg} is the one example
in this paper, through its denominator.

\paragraph{Ensemble universality.} It would also be interesting to extend
the results of this paper to more general ensembles of Hamiltonians; a
potentially direct generalization could be in the case of the ensemble dual
to $q$-deformed JT gravity. For Dyson indices $\mathsf b=1$ (GOE) and
$\mathsf b=4$ (GSE) the connected two-level correlator is replaced by the
corresponding cluster functions~\cite{Mehta:2004}, changing the SFF ramp
but not the genus-zero negativity, which uses only the smooth density of
states. Finally, our results may have some bearing on the question of
whether gravity should be thought of as being dual to one boundary theory
or an ensemble average over boundary theories.

\paragraph{Beyond JT.} Three natural extensions follow. (i) JT
supergravity: the disk density of states and the length-basis
wavefunctions are modified~\cite{Saad:2019lba,Mertens:2022irh}, and it
would be interesting to check whether the Gaussian-packet structure of the
HH state at high temperature persists. (ii) This discussion seems very
tailored to $0+1$-dimensional quantum systems, and it would be interesting
to explore whether gravity in higher dimensions could emerge from similar
considerations. (iii) Finite $\ee^{S_0}$: the Krylov-frame negativity of
the HH seed on a finite-$S_0$ JT spectrum is well defined where the length
basis is not, and is the natural object with which to ask how the
genus-zero value $1+\delta$ is lifted; the number of levels in the thermal
window, of order $\ee^{S_0}\ee^{2\pi^2/\beta}\beta^{-3/2}$, makes this
feasible only away from the high-temperature regime studied here. Finally,
it would be very interesting to explore these ideas in toy models for
holography such as the SYK or the DSSYK model.

\paragraph{Order of limits.} Presumably, continuum gravity emerges from an
appropriate coarse-graining in the large-$D$ limit. In the strict
genus-zero limit the negativity freezes at $1+\delta_\infty(\beta)$,
Eq.~\eqref{eq:deltainf}; at finite $\ee^{S_0}$ the fine-grained spectrum
is expected to lift it, to a value and on a time scale that are not fixed
by the disk. From this perspective, it is natural to think of the Krylov
phase space and the corresponding Hamiltonian as the dual
``gravitational'' theory, which seeks to implement the boundary quantum
dynamics on a classical computer to the best extent possible, supplemented
with some additional non-classical resource. The bulk gravitational
theory of AdS/CFT provides a semiclassical description of strongly
coupled, chaotic boundary dynamics at large $N$; the picture that gravity
should be read as a low-complexity effective theory has been advocated
in~\cite{Balasubramanian:2026wkv}. A claim that gravity is an efficient
classical simulation of boundary dynamics would require specifying the
class of operations, the target precision and the regularisation that
define the simulation problem in the continuum, which we have not done.
The conservative statement supported by this paper is that, in the toy
model setting of JT gravity, the length basis provides an unusually
Wigner-classical representation of the dynamics at genus zero,
Eq.~\eqref{eq:Ndisk}, with a negativity that is exponentially small in
$1/\beta$.

\begin{acknowledgments}
I thank Onkar Parrikar for guidance and numerous discussions, and the
members of the Department of Theoretical Physics at TIFR for a stimulating
environment. 
\end{acknowledgments}
\clearpage
\appendix

\section{Bessel asymptotics and the reflection phase}
\label{app:bessel}

In this Appendix, we present the details of the derivation
of~\eqref{eq:Kasym-general} and~\eqref{eq:stirlingphase} quoted in the main
text. Recall that we had the connection formula and the small-argument
expansion of the modified Bessel functions,
\beq
  \begin{split}
  K_{\ii\nu}(z)&=\frac{\pi}{2\ii\sinh(\pi\nu)}\big[I_{-\ii\nu}(z)-I_{\ii\nu}(z)\big],\\
  I_{\ii\nu}(z)&=\frac{(z/2)^{\ii\nu}}{\Gamma(1+\ii\nu)}\,\big(1+O(z^2)\big).
  \end{split}
  \label{eq:appA1}
\eeq
Writing $\Gamma(1+\ii\nu)=|\Gamma(1+\ii\nu)|\,\ee^{\ii\gamma}$ with
$\gamma=\arg\Gamma(1+\ii\nu)$, we have
\beq
  I_{\pm\ii\nu}(z)=\frac{\ee^{\pm\ii[\nu\log(z/2)-\gamma]}}{|\Gamma(1+\ii\nu)|}
  \big(1+O(z^2)\big),
  \label{eq:appA2}
\eeq
and using
$|\Gamma(1+\ii\nu)|^2=\pi\nu/\sinh(\pi\nu)$, which follows from the
reflection formula $\Gamma(1+\ii\nu)\Gamma(1-\ii\nu)=\pi\nu/\sinh\pi\nu$,
the difference in~\eqref{eq:appA1} becomes
\beq
  I_{-\ii\nu}(z)-I_{\ii\nu}(z)
  =\frac{-2\ii\,\sin\!\big[\nu\log\tfrac z2-\gamma\big]}
        {\sqrt{\pi\nu/\sinh\pi\nu}}\big(1+O(z^2)\big).
  \label{eq:appA3}
\eeq
Substituting~\eqref{eq:appA3} into~\eqref{eq:appA1}, the factors of $\ii$
and of $\sinh(\pi\nu)$ combine to give
\beq
  \begin{split}
  K_{\ii\nu}(z)
  ={}&-\sqrt{\frac{\pi}{\nu\sinh(\pi\nu)}}\,
  \sin\!\big[\nu\log\tfrac z2-\arg\Gamma(1+\ii\nu)\big]\\
  &\times\big(1+O(z^2)\big).
  \end{split}
  \label{eq:appA4}
\eeq
Now $\Gamma(1+\ii\nu)=\ii\nu\,\Gamma(\ii\nu)$, so
$\arg\Gamma(1+\ii\nu)=\tfrac\pi2+\arg\Gamma(\ii\nu)$, and
$-\sin(\theta-\tfrac\pi2)=\cos\theta$; also
$\nu\log\tfrac z2=-\nu\log\tfrac2z$. This leads to
\beq
  \begin{split}
  K_{\ii\nu}(z)
  &=\sqrt{\frac{\pi}{\nu\sinh(\pi\nu)}}\,
   \cos\!\big[\nu\log\tfrac2z+\arg\Gamma(\ii\nu)\big]\\
  &\quad\times\big(1+O(z^2)\big).
  \end{split}
  \label{eq:appA5}
\eeq
Setting $\nu=2s$ gives~\eqref{eq:Kasym-general}, and setting
$z=4\ee^{-x/2}$, i.e.\ $\log(2/z)=\tfrac x2-\log2$, gives~\eqref{eq:Kasym}
with $\phi_0(s)=2s\log2-\arg\Gamma(2\ii s)$.

For the Stirling phase, apply
$\log\Gamma(w)=(w-\tfrac12)\log w-w+\tfrac12\log2\pi+O(w^{-1})$ at $w=\ii y$,
$y>0$, where $\log(\ii y)=\log y+\ii\tfrac\pi2$:
\beq
  \begin{split}
  \mathrm{Im}\,\log\Gamma(\ii y)
  &=\mathrm{Im}\Big[(\ii y-\tfrac12)(\log y+\ii\tfrac\pi2)-\ii y\Big]+O(y^{-1})\\
  &=y\log y-y-\frac\pi4+O(y^{-1}).
  \end{split}
  \label{eq:appA6}
\eeq
(The next term of the Stirling series is $1/(12w)=-\ii/(12y)$, so the
remainder in~\eqref{eq:appA6} is $-1/(12y)$.) Setting $y=2s$ gives the first equation
of~\eqref{eq:stirlingphase}; then
\beq
  \begin{split}
  \phi_0(s)&=2s\log2-2s\log(2s)+2s+\tfrac\pi4+O(s^{-1})\\
  &=-2s\log s+2s+\tfrac\pi4+O(s^{-1}).
  \end{split}
  \label{eq:appA7}
\eeq
Differentiating the asymptotic series term by term gives
$\phi_0'(s)=-2\log s+O(s^{-2})$, $\phi_0''(s)=-2/s+O(s^{-3})$ and
$\phi_0'''(s)=2/s^2+O(s^{-4})$, which is~\eqref{eq:phi0derivs}; the phase
of~\eqref{eq:Kasym} is then
$sx-\phi_0(s)=s(x+2\log s-2)-\tfrac\pi4+O(s^{-1})$, i.e.\ with $E=s^2$,
$\sqrt E\,(x+\log E-2)-\tfrac\pi4$ (note the single $\log E$, not
$\log4E$).

\section{Normalisation of the Hartle--Hawking state}
\label{app:norm}

In this section, we will evaluate the norm of the state obtained in
equation~\eqref{eq:psi}, and fix its prefactor. Writing
$\psi_\beta=\mathcal C\int_0^\infty 2s\,\dd s\,\rho_0(s)\ee^{-\beta s^2/2}
K_{2\ii s}(4\ee^{-x/2})$,
\beq
  \begin{split}
  \langle\psi_\beta|\psi_\beta\rangle
  =|\mathcal C|^2\!\int_0^\infty\!\!\int_0^\infty\!
  &2s\,\dd s\,2s'\dd s'\,\rho_0(s)\rho_0(s')\\
  &\times\ee^{-\frac{\beta}{2}(s^2+s'^2)}\,
  \mathcal K(s,s'),
  \end{split}
  \label{eq:appB1}
\eeq
where
$\mathcal K(s,s')\equiv\int_{-\infty}^{\infty}\dd x\,
K_{2\ii s}(4\ee^{-x/2})K_{2\ii s'}(4\ee^{-x/2})$. We can simplify the
remaining integral in the following way: with $z=4\ee^{-x/2}$ we have
$\dd x=-2\,\dd z/z$, and the Kontorovich--Lebedev orthogonality relation
\beq
  \int_0^\infty K_{\ii\nu}(z)K_{\ii\nu'}(z)\,\frac{\dd z}{z}
  =\frac{\pi^2\,\delta(\nu-\nu')}{2\nu\sinh(\pi\nu)}
  \label{eq:appB2}
\eeq
with $\nu=2s$, $\nu'=2s'$ and $\delta(2s-2s')=\tfrac12\delta(s-s')$ gives
\beq
  \mathcal K(s,s')=\frac{\pi^2\,\delta(s-s')}{4s\,\sinh(2\pi s)}.
  \label{eq:appB3}
\eeq
Substituting
into~\eqref{eq:appB1}, the $s'$ integral collapses, and since
$\rho_0(s)=\sinh(2\pi s)$ one power of $\sinh(2\pi s)$ cancels:
\beq
  \langle\psi_\beta|\psi_\beta\rangle
  =|\mathcal C|^2\,\pi^2\!\int_0^\infty\! s\,\sinh(2\pi s)\,\ee^{-\beta s^2}\dd s
  =|\mathcal C|^2\,\frac{\pi^2}{2}\,Z_\beta ,
  \label{eq:appB4}
\eeq
by the definition~\eqref{eq:Zdef} of $Z_\beta$. Thus
$\langle\psi_\beta|\psi_\beta\rangle=1$ requires
$\mathcal C=\sqrt2/(\pi\sqrt{Z_\beta})$, which is the prefactor
in~\eqref{eq:psi}; the prefactor $1/(\pi\sqrt{Z_\beta})$ would give
$\langle\psi_\beta|\psi_\beta\rangle=\tfrac12$.

\section{The Hartle--Hawking packet at the turning point}
\label{app:packet}

We derive the leading semiclassical form~\eqref{eq:HHgauss} of the HH
wavefunction~\eqref{eq:psi} at $t=0$. Insert the integral representation
$K_{2\ii s}(z)=\int_0^\infty\dd v\,\ee^{-z\cosh v}\cos(2sv)$, valid for
$z>0$, and exchange the (absolutely convergent) integrals:
\beq
  \psi_\beta(x)=\frac{\sqrt2}{\pi\sqrt{Z_\beta}}\int_0^\infty\!\dd v\;
  \ee^{-z\cosh v}\,J(v),\qquad z=4\ee^{-x/2},
  \label{eq:appD1}
\eeq
with
\beq
  J(v)=\int_0^\infty\!\dd s\;2s\sinh(2\pi s)\,\ee^{-\beta s^2/2}\cos(2sv).
  \label{eq:appD2}
\eeq
The integrand of~\eqref{eq:appD2} is even in $s$, so $J$ is half the
integral over the real line; writing $\sinh(2\pi s)=\tfrac12(\ee^{2\pi s}-\ee^{-2\pi s})$
and $\cos(2sv)=\mathrm{Re}\,\ee^{2\ii sv}$, it reduces to the Gaussian
moments $\int_{-\infty}^{\infty}\dd s\,s\,\ee^{-\beta s^2/2+as}
=(a/\beta)\sqrt{2\pi/\beta}\,\ee^{a^2/2\beta}$ at $a=\pm2\pi+2\ii v$:
\beq
  \begin{split}
  J(v)&=\frac{2}{\beta}\sqrt{\frac{2\pi}{\beta}}\;\ee^{2\pi^2/\beta}\,
  \mathrm{Re}\Big[(\pi+\ii v)\,\ee^{-2v^2/\beta+4\pi\ii v/\beta}\Big]\\
  &=\frac{2}{\beta}\sqrt{\frac{2\pi}{\beta}}\;\ee^{2\pi^2/\beta}\,\ee^{-2v^2/\beta}
  \Big[\pi\cos\frac{4\pi v}{\beta}-v\sin\frac{4\pi v}{\beta}\Big].
  \end{split}
  \label{eq:appD3}
\eeq
Since $J$ is even in $v$, \eqref{eq:appD1} becomes~\eqref{eq:psi-v}.

Now set $x=x_{\rm turn}(s_*)+\Delta$ with $x_{\rm turn}(s_*)=2\log(2\beta/\pi)$,
so that $z=(2\pi/\beta)\,\ee^{-\Delta/2}$ and the exponent
of~\eqref{eq:psi-v} is $-G(v)/\beta$ with
\beq
  G(v)=2\pi\,\ee^{-\Delta/2}\cosh v+2v^2-4\pi\ii v .
  \label{eq:appD4}
\eeq
For $\beta\ll1$ the integral is dominated by the saddle point $G'(v)=0$,
i.e.\ $2\pi\ee^{-\Delta/2}\sinh v+4v-4\pi\ii=0$. Writing $v=\ii\theta$ this
is real,
\beq
  \sin\theta=\frac{2}{\pi}\,(\pi-\theta)\,\ee^{\Delta/2},
  \label{eq:appD5}
\eeq
with a unique solution $\theta(\Delta)\in(0,\pi)$ for
$\Delta<2\log(\pi/2)$, since $\sin\theta/(\pi-\theta)$ increases
monotonically from $0$ to $1$ on $(0,\pi)$; it has $\theta(0)=\pi/2$. [The
endpoint $\theta=\pi$ solves~\eqref{eq:appD5} for every $\Delta$, but it
is the saddle at $v=\ii\pi$, where $\cosh v=-1$ and the integration
contour cannot be deformed; it does not contribute.] At the saddle both the exponent and the prefactor
are real,
\beq
  \begin{split}
  F(\Delta)&\equiv G(\ii\theta)=2\pi\ee^{-\Delta/2}\cos\theta-2\theta^2+4\pi\theta,\\
  \pi+\ii v_*&=\pi-\theta ,
  \end{split}
  \label{eq:appD6}
\eeq
so $\psi_\beta$ is real and positive near the peak, as it must be. Since
$\partial_\theta G(\ii\theta)=0$ at the saddle, $F'(\Delta)=\partial_\Delta G
=-\pi\ee^{-\Delta/2}\cos\theta$, which vanishes exactly at $\theta=\pi/2$,
i.e.\ at $\Delta=0$: the exponent is extremal at the classical turning
point. Differentiating~\eqref{eq:appD5} at $\Delta=0$ gives
$\theta'(0)=\pi/4$, and hence
\beq
  F(0)=\frac{3\pi^2}{2},\quad F'(0)=0,\quad
  F''(0)=\pi\,\theta'(0)=\frac{\pi^2}{4}>0 ,
  \label{eq:appD7}
\eeq
so $\Delta=0$ is the minimum of $F$ and the peak of $|\psi_\beta|^2$. The
Gaussian fluctuation integral about the saddle uses
$G''(\ii\theta)=2\pi\ee^{-\Delta/2}\cos\theta+4$, equal to $4$ at
$\Delta=0$, so that $\int\dd w\,\ee^{-G''w^2/2\beta}=\sqrt{\pi\beta/2}$.
Assembling the prefactor of~\eqref{eq:psi-v} with
$Z_\beta=\pi^{3/2}\beta^{-3/2}\ee^{\pi^2/\beta}$ from~\eqref{eq:Zdef},
\beq
  \psi_\beta(x_{\rm turn})\simeq
  \frac{2\,\ee^{2\pi^2/\beta}}{\sqrt{\pi Z_\beta}\,\beta^{3/2}}\cdot\frac\pi2\cdot
  \ee^{-3\pi^2/2\beta}\sqrt{\frac{\pi\beta}{2}}
  =\Big(\frac{\pi}{4\beta}\Big)^{1/4}:
  \label{eq:appD8}
\eeq
the exponentials cancel identically,
$2\pi^2-\tfrac32\pi^2-\tfrac12\pi^2=0$, and the prefactor is precisely the
peak value of a unit-normalised Gaussian of variance
$\sigma_x^2=\beta/(2F''(0))=2\beta/\pi^2$. This is~\eqref{eq:HHgauss}. The
cancellation of the exponentials and the emergence of the correct
normalisation are the consistency check between the saddle-point
evaluation and the exact norm of Appendix~\ref{app:norm}.

The corrections to~\eqref{eq:HHgauss} come from the cubic term
$F'''(0)\Delta^3/6\beta$ and from the $\Delta$ dependence of the prefactor
$(\pi-\theta)/\sqrt{G''}$; both are odd in $\Delta$ and of relative size
$O(\Delta)=O(\sqrt\beta)$ across the width of the packet. The moments
quoted in~\eqref{eq:HHmoments} follow from the Gaussian:
$\sigma_x^2=2\beta/\pi^2$ and, since $\psi_\beta$ is real,
$\langle p\rangle=0$ and $\langle p^2\rangle=1/(4\sigma_x^2)=\pi^2/(8\beta)$.
Consistently, $\langle E\rangle_\beta-\langle p^2\rangle=\langle4\ee^{-x}\rangle$
is $s_*^2+O(1/\beta)$: at $t=0$ the energy of the packet is potential
energy.

\section{The Laguerre reference chain}
\label{app:laguerre}

In this Appendix, we present the details of the derivation
of~\eqref{eq:laguerre} quoted in the main text. Set $\beta=1$ (restored by
$E\to\beta E$, $t\to t/\beta$). Recall that the generalised Laguerre
polynomials $L_n^{(\alpha)}$ are orthogonal with respect to the weight
$x^\alpha\ee^{-x}$ on $[0,\infty)$, with
$\int_0^\infty x^\alpha\ee^{-x}L_n^{(\alpha)}L_m^{(\alpha)}\dd x
=\frac{\Gamma(n+\alpha+1)}{n!}\delta_{nm}$, and satisfy the three-term
recurrence
\beq
  x\,L_n^{(\alpha)}=-(n+1)L_{n+1}^{(\alpha)}+(2n+\alpha+1)L_n^{(\alpha)}
  -(n+\alpha)L_{n-1}^{(\alpha)} .
  \label{eq:appE1}
\eeq
The orthonormal polynomials of the normalised weight
$x^\alpha\ee^{-x}/\Gamma(\alpha+1)$ are therefore
$\hat p_n=\sqrt{n!\,\Gamma(\alpha+1)/\Gamma(n+\alpha+1)}\,L_n^{(\alpha)}$,
and rewriting~\eqref{eq:appE1} in terms of $\hat p_n$ gives
$x\hat p_n=b_{n+1}\hat p_{n+1}+a_n\hat p_n+b_n\hat p_{n-1}$ (up to the
sign of $\hat p_{n+1}$, which is a choice of phase) with
\beq
  a_n=2n+\alpha+1,\qquad b_n=\sqrt{n(n+\alpha)} ,
  \label{eq:appE2}
\eeq
since $(n+1)\sqrt{\tfrac{n!\,\Gamma(n+\alpha+2)}{(n+1)!\,\Gamma(n+\alpha+1)}}
=\sqrt{(n+1)(n+1+\alpha)}$. These are the Lanczos coefficients of the
chain, in the same sense as in Sec.~\ref{sec:spread}: the Krylov basis
generated by multiplication by $x$ acting on the constant function is the
orthonormal polynomial basis, and $a_n$, $b_n$ are its recursion
coefficients.

For the amplitudes $\varphi_n(t)=\langle K_n|\ee^{-\ii Ht}|K_0\rangle
=\int_0^\infty\frac{x^\alpha\ee^{-x}}{\Gamma(\alpha+1)}\,\hat p_n(x)\,\ee^{-\ii xt}\dd x$,
we need the classical integral
\beq
  \begin{split}
  &\int_0^\infty\!\dd x\,x^{\alpha}\ee^{-(1+\ii t)x}L_n^{(\alpha)}(x)\\
  &\qquad=\frac{\Gamma(n+\alpha+1)}{n!}
   \frac{(\ii t)^n}{(1+\ii t)^{n+\alpha+1}},
  \end{split}
  \label{eq:appE3}
\eeq
To see this, multiply the generating function
$\sum_nL_n^{(\alpha)}(x)z^n=(1-z)^{-\alpha-1}\ee^{xz/(z-1)}$ by
$x^\alpha\ee^{-x(1+\ii t)}$ and integrate:
\beq
  \begin{split}
  &\sum_n z^n\!\int_0^\infty\! x^\alpha\ee^{-x(1+\ii t)}L_n^{(\alpha)}\dd x\\
  &\quad=(1-z)^{-\alpha-1}\!\int_0^\infty\! x^\alpha\ee^{-x\left(1+\ii t+\frac{z}{1-z}\right)}\dd x\\
  &\quad=\frac{\Gamma(\alpha+1)\,(1-z)^{-\alpha-1}}{\big(1+\ii t+\frac{z}{1-z}\big)^{\alpha+1}}
  =\frac{\Gamma(\alpha+1)}{\big(1+\ii t-\ii tz\big)^{\alpha+1}},
  \end{split}
  \label{eq:appE4}
\eeq
and expanding $(1+\ii t)^{-\alpha-1}\big(1-\tfrac{\ii t}{1+\ii t}z\big)^{-\alpha-1}$
in powers of $z$ with the binomial series
$(1-w)^{-\alpha-1}=\sum_n\binom{n+\alpha}{n}w^n$ gives the coefficient
of $z^n$ as $\Gamma(\alpha+1)\binom{n+\alpha}{n}(\ii t)^n/(1+\ii t)^{n+\alpha+1}$,
which is~\eqref{eq:appE3} since
$\Gamma(\alpha+1)\binom{n+\alpha}{n}=\Gamma(n+\alpha+1)/n!$. Including the normalisation of $\hat p_n$,
\beq
  |\varphi_n(t)|^2
  =\binom{n+\alpha}{n}\frac{t^{2n}}{(1+t^2)^{\,n+\alpha+1}},
  \label{eq:appE5}
\eeq
a negative-binomial distribution in $n$ with success parameter
$p=t^2/(1+t^2)$: indeed
$\sum_n\binom{n+\alpha}{n}p^n(1-p)^{\alpha+1}=1$ is the binomial series
again, and its mean is
\beq
  \sum_n n\binom{n+\alpha}{n}p^n(1-p)^{\alpha+1}
  =(\alpha+1)\frac{p}{1-p}=(\alpha+1)\,t^2 ,
  \label{eq:appE6}
\eeq
which is~\eqref{eq:laguerre}. Finally $b_1^2=\alpha+1$ by~\eqref{eq:appE2},
which closes the identity $\Cplx_L=b_1^2t^2$ for all $t$.

\section{Short-time expansion of the spread complexity from cumulants}
\label{app:cumulants}

Let $\mu_k=\langle K_0|H^k|K_0\rangle$ be the moments of the seed and
$\kappa_r$ the corresponding cumulants; for the thermal seed,
$\kappa_r=(-1)^r\partial_\beta^r\log Z_\beta$. Write $\tilde H=H-a_0$ with
$a_0=\mu_1=\kappa_1$, so that $\langle\tilde H^2\rangle=\kappa_2$,
$\langle\tilde H^3\rangle=\kappa_3$ and
$\langle\tilde H^4\rangle=\kappa_4+3\kappa_2^2$. The Lanczos recursion
gives $b_1^2=\langle\tilde H^2\rangle=\kappa_2$ and
$|K_1\rangle=\tilde H|K_0\rangle/b_1$, hence
\beq
  a_1=\langle K_1|H|K_1\rangle
  =\frac{\langle\tilde H^3\rangle}{\kappa_2}+a_0
  \;\Rightarrow\;
  a_1-a_0=\frac{\kappa_3}{\kappa_2}.
  \label{eq:appF1}
\eeq
From $(H-a_1)|K_1\rangle=b_2|K_2\rangle+b_1|K_0\rangle$,
$b_2^2=\langle K_1|(H-a_1)^2|K_1\rangle-b_1^2$, and expanding
$(H-a_1)^2=\tilde H^2-2(a_1-a_0)\tilde H+(a_1-a_0)^2$ inside
$\langle K_1|\cdots|K_1\rangle=\langle K_0|\tilde H\cdots\tilde H|K_0\rangle/\kappa_2$,
\beq
  b_2^2=\frac{\kappa_4}{\kappa_2}+2\kappa_2-\frac{\kappa_3^2}{\kappa_2^2}.
  \label{eq:appF2}
\eeq
Now expand the Krylov amplitudes
$\varphi_n(t)=\langle K_n|\ee^{-\ii Ht}|K_0\rangle$ to the order needed for
$\Cplx=\sum_nn|\varphi_n|^2$ through $t^4$. Using
$\langle K_1|H^2|K_0\rangle=b_1(a_0+a_1)$,
$\langle K_1|H^3|K_0\rangle=b_1(a_0^2+a_0a_1+a_1^2+b_1^2+b_2^2)$ and
$\langle K_2|H^2|K_0\rangle=b_1b_2$,
\beq
  \begin{split}
  \varphi_1&=-\ii b_1t-\frac{t^2}{2}b_1(a_0+a_1)\\
  &\quad+\frac{\ii t^3}{6}b_1\big(a_0^2+a_0a_1+a_1^2+b_1^2+b_2^2\big)+O(t^4),\\
  \varphi_2&=-\frac{t^2}{2}b_1b_2+O(t^3).
  \end{split}
  \label{eq:appF3}
\eeq
In $|\varphi_1|^2$ the $O(t^3)$ cross term vanishes, and at $O(t^4)$ the
square of the $t^2$ term combines with the cross term between the $t$ and
$t^3$ terms:
\beq
  \begin{split}
  \Cplx(t)&=b_1^2t^2+b_1^2\Big[\frac{(a_0+a_1)^2}{4}\\
  &\qquad-\frac{a_0^2+a_0a_1+a_1^2+b_1^2+b_2^2}{3}+\frac{b_2^2}{2}\Big]t^4+O(t^6)\\
  &=b_1^2t^2+b_1^2\Big[-\frac{(a_1-a_0)^2}{12}-\frac{b_1^2}{3}+\frac{b_2^2}{6}\Big]t^4\\
  &\qquad+O(t^6),
  \end{split}
  \label{eq:appF4}
\eeq
where the second line uses
$\tfrac14(a_0+a_1)^2-\tfrac13(a_0^2+a_0a_1+a_1^2)=-\tfrac1{12}(a_1-a_0)^2$.
Substituting~\eqref{eq:appF1} and~\eqref{eq:appF2}, the $\kappa_2$ terms
cancel and
\beq
  \Cplx(t)=\kappa_2t^2+\Big(\frac{\kappa_4}{6}-\frac{\kappa_3^2}{4\kappa_2}\Big)t^4+O(t^6),
  \label{eq:appF5}
\eeq
which is~\eqref{eq:C4general}. For the Gamma (Laguerre) measure
$E^\alpha\ee^{-\beta E}$ the cumulants are $\kappa_r=(\alpha+1)(r-1)!/\beta^r$,
so $\kappa_4/6=(\alpha+1)/\beta^4=\kappa_3^2/4\kappa_2$ and the $t^4$ term
vanishes, as it must for the exactly quadratic law~\eqref{eq:appE6}. For
the JT measure, $\log Z_\beta=\text{const}-\tfrac32\log\beta+\pi^2/\beta$
gives
\beq
  \begin{split}
  \kappa_2&=\frac{4\pi^2+3\beta}{2\beta^3},\qquad
  \kappa_3=\frac{3(2\pi^2+\beta)}{\beta^4},\\
  \kappa_4&=\frac{3(8\pi^2+3\beta)}{\beta^5},
  \end{split}
  \label{eq:appF6}
\eeq
and
\beq
  \begin{split}
  \frac{\kappa_4}{6}-\frac{\kappa_3^2}{4\kappa_2}
  &=\frac{(3\beta+8\pi^2)(4\pi^2+3\beta)-9(\beta+2\pi^2)^2}{2\beta^5(4\pi^2+3\beta)}\\
  &=-\frac{2\pi^4}{\beta^5(4\pi^2+3\beta)},
  \end{split}
  \label{eq:appF7}
\eeq
which is~\eqref{eq:C4}.
\newpage
\bibliographystyle{apsrev4-2}
\bibliography{references}

\end{document}